\documentclass[a4paper]{article}

\usepackage{environ}
\usepackage{authblk}

\NewEnviron{ShrinkMath}{%
    \begin{equation}
    \scalebox{0.9}{$\BODY$}
    \end{equation}
    }

\usepackage{tcolorbox}
\usepackage{multicol}
\usepackage{charter}
\usepackage{helvet}
\usepackage[T1]{fontenc}
\usepackage[a4paper]{geometry}
\usepackage{xcolor}
\usepackage{mathrsfs}
\usepackage{mathtools}
\usepackage{multirow}
\usepackage{amsmath}
\usepackage{tabularx}
\usepackage{amssymb}
\usepackage{graphicx}
\usepackage{setspace}
\usepackage{caption}
\usepackage{algorithm}
\usepackage{algpseudocode}
\usepackage[unicode=true,
 bookmarks=false,
 breaklinks=false,pdfborder={0 0 1},backref=false,colorlinks=false]
 {hyperref}
\usepackage{enumitem}
\makeatletter

\usepackage{xcolor}
\usepackage{lscape}

\usepackage{savesym}
\numberwithin{equation}{section}
\usepackage{caption}
\usepackage{subcaption}
\usepackage{verbatim}
\usepackage{epsfig}
\usepackage{asymptote}
\graphicspath{{Figures/}} 

\savesymbol{amalg}
\usepackage{mathabx}
\restoresymbol{MAT}{amalg}

\savesymbol{ulcorner}
\savesymbol{urcorner}
\savesymbol{llcorner}
\savesymbol{lrcorner}
\usepackage{amsfonts}
\restoresymbol{AMS}{ulcorner}
\restoresymbol{AMS}{urcorner}
\restoresymbol{AMS}{llcorner}
\restoresymbol{AMS}{lrcorner}
\usepackage[bbgreekl]{mathbbol}
\DeclareSymbolFontAlphabet{\mathbbm}{bbold}
\DeclareSymbolFontAlphabet{\mathbb}{AMSb}

\usepackage{amsthm}
\usepackage{cases}
\usepackage{algorithm}
\usepackage{algpseudocode}
\usepackage{pict2e}
\usepackage{mwe}
\usepackage{footmisc}
\usepackage[cal=boondox,scr=boondoxo]{mathalfa}
\usepackage{multirow}
\usepackage{array}
\usepackage{colortbl}

\usepackage{soul}
\setstcolor{red}
\usepackage{todonotes}

\usepackage{enumitem}

\usepackage{url}
\usepackage{mwe}

    \AfterLastShipout{\immediate\write\@mainaux{\relax}}%

\newif\ifHNNdraft
\HNNdrafttrue

\PassOptionsToPackage{normalem}{ulem}
\usepackage{ulem}

\newcommand\footnoteref[1]{\protected@xdef\@thefnmark{\ref{#1}}\@footnotemark}

\usepackage{nomencl}
\makenomenclature

\renewcommand{\u}[1]{\boldsymbol{#1}}
\newcommand{\ou}[1]{\overline{\u{#1}}}

\newcommand{\pr}[1]{\left( #1 \right)}

\newcommand{\ag}[1]{\left[ #1 \right]}
\newcommand{\physC}{\boldsymbol{\mathscr{C}}}

\newcommand{\set}[2]{\left\{#1\, \big|\, #2\right\}}
\newcommand{\norm}[1]{\lVert #1\rVert}
\newcommand{\abs}[1]{\left\lvert #1\right\rvert}

\newcommand{\vast}{\bBigg@{4}}
\newcommand{\Vast}{\bBigg@{5}}

\DeclareMathAlphabet{\mathpzc}{OT1}{pzc}{m}{it}

\makeatletter
\def\@gobbleappendixname#1\csname thesection\endcsname{\Alph{section}.\arabic{subsection}}
\g@addto@macro{\appendix}{\renewcommand{\p@subsection}{\@gobbleappendixname}}
\makeatother

\newcommand*{\inlineequation}[2][]{%
  \begingroup
    \refstepcounter{equation}%
    \ifx\\#1\\%
    \else
      \label{#1}%
    \fi
    \relpenalty=10000 %
    \binoppenalty=10000 %
    \ensuremath{%
      #2%
    }%
    ~\@eqnnum
  \endgroup
}

\makeatother
\makeatletter
\newcommand*\@dblLabelI {}
\newcommand*\@dblLabelII {}
\newcommand*\@dblequationAux {}

\def\@dblequationAux #1,#2,%
    {\def\@dblLabelI{\label{#1}}\def\@dblLabelII{\label{#2}}}

\newcommand*{\doubleequation}[3][]{%
    \par\vskip\abovedisplayskip\noindent
    \if\relax\detokenize{#1}\relax
       \let\@dblLabelI\@empty
       \let\@dblLabelII\@empty
    \else 
       \@dblequationAux #1,%
    \fi
    \makebox[0.5\linewidth-1.5em]{%
     \hspace{\stretch2}%
     \makebox[0pt]{$\displaystyle #2$}%
     \hspace{\stretch1}%
    }%
    \makebox[0.5\linewidth-1.5em]{%
     \hspace{\stretch1}%
     \makebox[0pt]{$\displaystyle #3$}%
     \hspace{\stretch2}%
    }%
    \makebox[3em][r]{(%
  \refstepcounter{equation}\theequation\@dblLabelI,
  \refstepcounter{equation}\theequation\@dblLabelII)}%
  \par\vskip\belowdisplayskip
}
\makeatother

\begin{document}

\title{A finite rotation, small strain  2D  elastic head model, with applications in mild traumatic brain injury}
\author[1]{Yang Wan}
\author[1]{Wenqiang Fang}
\author[2]{Rika Wright Carlsen}
\author[1,*]{Haneesh Kesari}
\affil[1]{School of Engineering, Brown University, Providence, RI 02912, USA}
\affil[2]{Department of Engineering, Robert Morris University, Moon Township, PA 15108, USA}
\affil[*]{Corresponding author, haneesh\_kesari@brown.edu}

\maketitle

\bibliographystyle{elsarticle-num}

\begin{abstract}
Rotational head motions have been shown to play a key role in traumatic brain injury.  There is great interest in developing methods to rapidly predict brain tissue strains and strain rates resulting from rotational head motions to estimate brain injury risk and to guide the design of protective equipment.  Idealized continuum mechanics based head models provide an attractive approach for rapidly estimating brain strains and strain rates.  These models are capable of capturing the wave dynamics and transient response of the brain while being significantly easier and faster to apply compared to more sophisticated and detailed finite element head models.  In this work, we present a new idealized continuum mechanics based head model that accounts for the head's finite rotation, which is an improvement upon prior models that have been based on a small rotation assumption.  Despite the simplicity of the model, we show that the proposed 2D elastic finite rotation head model predicts comparable strains to a more detailed finite element head model, demonstrating the potential usefulness of the model in rapidly estimating brain injury risk. This newly proposed model can serve as a basis for introducing finite rotations into more sophisticated head models in the future.
\end{abstract}



\section{Introduction}
\label{sec:intro}
Mild traumatic brain injury (mTBI), also referred to as concussion, is a common injury for both civilians and warfighters. Despite efforts to prevent, diagnose, and treat TBI more effectively, it has remained a persistent problem \cite{maas2022traumatic}.  A considerable number of people in the civilian and military populations continue to experience mTBI, despite the availability of several mTBI mitigation interventions and protocols \cite{maas2022traumatic,howard2022association, phipps2020characteristics}.  Therefore, there is an urgent need for accurate and usable \textit{tools or methodologies that can assess injury risk of a mechanically traumatic event}, such as in falls, accidental or intentional blunt trauma to the head, or vehicular accidents.  These tools and methodologies are necessary for the effective design of devices, materials, and protocols that can mitigate the incidence of mTBI.

Several approaches have been put forward for assessing the risk of traumatic brain injury, including \textit{(i)} empirical injury criteria, \textit{(ii)} computational mechanics (CM) based injury criteria, and \textit{(iii)} machine learning (ML) based injury criteria.  Examples of empirically derived injury criteria include the Head Injury Criterion (HIC), which estimates injury risk based on the measured linear acceleration of the head, and the  Brain Injury Criterion (BrIC), which estimates injury risk based on the measured angular velocity of the head \cite{versace1971review, takhounts2013development, greenwald2008head}.   Empirical injury criteria are popular given their ease of use.  They are easy to understand and apply since the calculations involve simple data processing and evaluation of simple algebraic mathematical expressions.  They also have low to no computational expense and do not require detailed personalized information (e.g., head magnetic resonance images (MRI)).  A major drawback of empirically based injury criteria is that they are overly simple, i.e., they do not include important physical quantities that are known to affect the brain tissue strains and strain rates.  Both brain strains and strain rates have been shown to play an important role in deformation-induced neural injury, which is a common injury mechanism in mTBI \cite{hajiaghamemar2021multi, bar2016strain}.  As a result, the effectiveness of these empirical injury criteria in predicting the risk of mTBI remains an open question.

In contrast to empirical injury criteria, computational mechanics (CM) based injury criteria attempt to account for all relevant physics that lead to injury in a mTBI event.  Computational mechanics based models of mTBI are implicitly based on the hypothesis that the risk of injury of a traumatic event is correlated with some measure of brain tissue strains and strain rates.   The application of CM based injury criteria involves the numerical simulation of head motion of a traumatic event of interest followed by a comparison of the predicted strain ($\epsilon$) and strain rate ($\dot{\epsilon}$) from the model with critical values for injury.  Some measures of strain that are widely used in CM based injury criteria are the peak \textit{maximum principal strain} (MPS), \textit{maximum axonal strain} (MAS) or \textit{tract-oriented strain}, and the \textit{cumulative strain damage measure} (CSDM) \cite{takhounts2003development, Carlsen2021, zhan2020prediction, giordano2017anisotropic, garimella2019embedded, zhou2021toward}.  Determining  the measures of strain and strain rate that are most pertinent for mTBI and determining their critical values, i.e., the values at which the risk of injury becomes significant, is an active area of research \cite{bar2016strain, estrada2021neural, hajiaghamemar2021multi}.
A range of values have been proposed for various  critical strain measures, with the majority of estimates (maximum principal Green-Lagrange strain \cite{bain2000tissue}, MAS \cite{hajiaghamemar2020head}, and CSDM \cite{takhounts2003development}) falling between $10\%$ and $25\%$. There have been relatively fewer proposals for the critical values of strain rate measures. A recent study proposed a critical \textit{maximum axonal strain rate} (MASR) value of $40-90~{\rm s}^{-1}$   for traumatic axonal injury \cite{hajiaghamemar2021multi}. With time, both the measure used to predict injury as well as the accuracy of the critical values for that measure are expected to improve.

The effectiveness of CM based injury criteria will depend on how easily the strains and strain rates in the computational head model can be calculated and how close those calculated values are to their respective values during the real traumatic event.  The accuracy of the strain and strain rate estimates are dependent on several factors, such as the level of incorporated anatomical detail in the model, the numerical methods used, and the quality of the computational mesh.  A large number of computational head models have been developed, and they range from high resolution models that take into account the anisotropy of white matter and the detailed geometry of the brain to more simplified models that incorporate fewer anatomical details and have a lower spatial resolution \cite{giudice2019analytical, madhukar2019finite, dixit2017review}.  The computational time involved in simulating a head impact event is directly related to the mesh resolution.  The person-hours involved in preparing the mesh is also directly dependent on the quality and accuracy of the mesh.  For these reasons, getting accurate calculations of strains and strain rates in CM based head models is currently very computationally expensive, and from a  person-hours point of view, very time consuming, complex, and tedious.  In the future, the time to create these models is expected to decrease with the development of robust algorithms for automatically segmenting and meshing the brain directly from medical imaging data \cite{li2021subject, giudice2020image}. Even with these improvements, the computational expense of simulating traumatic events with high resolution, subject-specific models can still be prohibitive for some applications, given that these simulations can take hours or days to run even with the use of high performance computers.

The need to rapidly predict brain strains and strain rates to obtain real time estimates of mTBI risk has been a factor in the development of machine learning (ML) based methods as an alternative to CM based head models.  In this approach, ML algorithms are trained using CM based models, and once trained, the ML algorithms are applied to rapidly predict the risk of injury for a given head impact event.
Machine learning based methods have shown promise in predicting brain strain directly from head kinematic data \cite{upadhyay2022data, zhan2021rapid, wu2022real}.
However, ML derived injury measures still have some limitations. For example, there is a large cost upfront to develop ML based injury criteria.  Furthermore, subject-specific CM based models now take personal details, such as head size and shape, into account whereas ML based injury criteria usually do not. Machine learning based injury criteria can, of course, always be updated so that  the details of an individual are taken into account. This can be done, for example, by generating new data from CM based head models that have been updated by taking the individual's details into account and then training the ML based head model on that data. However, if such an approach is followed, then the development of ML based injury criteria will end up being even more expensive than the development of CM based injury criteria.

An alternative to empirically based, CM based, and ML based injury criteria are idealized continuum mechanics (ICM) based injury criteria. Idealized continuum mechanics based injury criteria are similar to CM based injury criteria in every way except that in them the head models used to estimate brain strains and strain rates are far more idealized than those used in CM criteria.
We will call the head model in an ICM based injury criteria an ICM based head model.
In most ICM based head models, the brain is simplified as a homogeneous, isotropic, incompressible, linear elastic or viscoelastic solid and the skull as a rigid solid; the geometry of the brain is modelled as a sphere or cylinder and the corresponding skull is modelled as a spherical shell or a hollow cylinder \cite{Ljung1975,Margulies1989,Massouros2014,Massouros2008,Massouros2005,Bayly2008,Lee1970,Firoozbakhsh1975,Bycroft1973,Liu1974,Liu1973,Misra1984,Liu1975,Christensen1964,Chandran1975,Cotter2002}. Due to that simplicity, there is no need for a computational mesh, and there is a large reduction in computational expense.

\paragraph{2D ICM models} There exist several ICM based head models in which the head is modeled as a  cylinder \cite{Ljung1975,Margulies1989,Massouros2014,Massouros2008,Massouros2005,Bayly2008}.
In all the 2D models we surveyed the head was subjected to a rotational motion with no translations being involved. The rotations take place about a fixed central axis, which is an axis that is fixed in space w.r.t. time and is initially the cylinder's central axis. In the context of ICM models ``2D''  means that the displacements in the direction of the central axis are assumed to vanish, and that the other displacement components are assumed to not vary in the direction of the central axis.

In the work of Margulies and Thibault \cite{Margulies1989} the brain is modeled as an incompressible, homogeneous Kelvin-Voigt viscoelastic material, and the skull is  subjected to a sudden time varying rotation.
The rotation  can be fairly arbitrary as long as the corresponding angular acceleration
 can be represented using a Fourier series.
 In the work of Massouros \emph{et al.} \cite{Massouros2014,Massouros2008,Massouros2005} the brain is modeled as a Maxwell viscoelastic material, and the skull's rotation angle is a sinusoidal function of time.
In the work of Bayly \emph{et al.} \cite{Bayly2008} the brain is modeled as a three-parameter linear viscoelastic material.
Initially the brain and the skull rotate with the same constant angular velocity, and the brain is curtailed from experiencing any deformation.
The head is then loaded by subjecting the skull to an angular deceleration pulse, which has the shape of a half-sine pulse, and the brain is allowed to experience deformation.
Bayly \textit{et al.} solve their model analytically as well as numerically using finite element methods.
They additionally validate their model by comparing its  predicted strain fields with the strain fields they measure in a gelatin based experiment that was set up to closely resemble their ICM based head model.

\paragraph{3D ICM models} There also exist several ICM based head models in which the brain is modeled as a sphere \cite{Christensen1964,Chandran1975,Ljung1975,Lee1970,Liu1973,Bycroft1973,Liu1974,Liu1975,Firoozbakhsh1975,Misra1984,Cotter2002}. Leaving \cite{Christensen1964, Chandran1975}, in all other models, the head is not subjected to any translatory motion; it is only subjected to a rotational motion about a fixed central axis, i.e., an axis that is fixed in space with time and that initially passes through the sphere's center.

In the work of Christensen and Gottenberg \cite{Christensen1964}, the brain is modeled as a general linear viscoelastic material, and the skull is first subjected to a rotational motion about a fixed central axis and then later to a translatory motion.
In the work of Chandran \emph{et al.} \cite{Chandran1975}, the brain is modeled as a linear elastic material, and the skull is subjected to purely translatory motion.

In the work of Lee and Advani \cite{Lee1970}, the brain is first modeled as a linear elastic material and then later as an arbitrary linear  viscoelastic solid; the skull is subjected to a time varying rotation. The function that maps time to the rotation angle is such that its second  derivative is a step function.
Firoozbakhsh and DeSilva \cite{Firoozbakhsh1975} and Bycroft \cite{Bycroft1973} model the brain as a general linear viscoelastic solid.  In the work of Firoozbakhsh and DeSilva, the function that maps time to the rotation angle can be fairly arbitrary, whereas in the work of Bycroft the function  is a half sine pulse.
Liu \emph{et al.} \cite{Liu1975} model the brain
first as a linear elastic material and then
as a linear Kelvin viscoelastic material; the skull is subjected to fairly arbitrary rotations.
Liu \emph{et al.} solved their ICM model using finite-difference methods.
For the case of the brain being linear elastic,
 Liu \emph{et al.}'s results \cite{Liu1975} match those of Liu and Chandran \cite{Liu1973}. The work of Liu and Chandran \cite{Liu1973} is cited often in the context of 3D ICM models. However, we were not able to access this publication. From the comments regarding it in \cite{Liu1975,Liu1974,Misra1984}, we know that in this work the brain is modeled as a linear elastic solid and the skull is subjected to fairly arbitrary rotations.

The  published ICM models that we surveyed  are  capable of capturing the wave dynamics in the brain due to sudden head motion, and provide a good first order approximation for the peak strains and strain rates in the brain. Despite the valuable preliminary physical insight that they provide, all the surveyed ICM models have a  key limitation. Their  derivation is based on the implicit assumption that the head rotations are small. This fact can be gleaned by noting that in the surveyed ICM models, the authors do not make a distinction between the reference and the deformed configurations. The head rotations in most mechanically traumatic events, however, are far from being small. In most mechanically traumatic events, such as vehicle crashes, sports injuries and maneuvers, martial arts, etc., the head rotations quite routinely exceed $45^{\circ}$, and in some cases even exceed  $90^{\circ}$ \cite{viano2007concussion, hernandez2019voluntary, li2018influence, yan2018investigation}.

In this paper, we present a new 2D ICM based head model that accounts for the head's finite rotations (see \S\ref{sec:governequ}).  Like all the surveyed ICM head models, the strains in it are still small; and like all the surveyed 2D ICM head models the loading in it is purely rotational, with no translations.
We refer to this new ICM based head model as the finite rotations head model.
On accounting for the head's finite rotations and distinguishing between the reference and deformed configurations, the equations (see \eqref{eq:BVPhatu2}) governing the motion of the brain in it come out to be quite different from those arrived at in the surveyed ICM models (see \eqref{eq:u2CBVPLinearized}); as we mentioned previously, the surveyed ICM models do not account for the head's finite rotations.
Using our model, we found that the estimates for various stress and strain measures, such as maximum principal logarithmic strain, can contain quite significant errors if the head's finite rotations are not taken into account (see \S\ref{sec:comMPLS}).  This, however, is not to say that our ICM model is better than the surveyed ICM models. In fact, our model is simpler than some of the surveyed ICM models in some aspects. For example,  it is 2D in nature, and takes the brain to be a perfectly elastic solid. Thus, our model's key contribution in the context of surveyed ICM models is in that it highlights the important role played by the finiteness of the head's  rotations in dictating the internal brain strains and strain rates. We believe that the presented finite rotations head model will prove useful for introducing finite rotations into more sophisticated head models, such as those that take the head's 3D nature and the brain's viscoelastic behavior into account.
%




This paper is organized as follows.  In \S\ref{sec:preliminaries}, we present the mathematical and mechanics preliminaries needed for the development of our finite rotations head model. In \S\ref{sec:governequ}, we present the equations that govern the mechanics of the finite rotations head model. In \S\ref{sec:anasol}, we provide a semi-analytical solution for the governing equations and derive expressions for the strains and strain rates in the finite rotations head model. In \S\ref{sec:femcompare}, we compare the strains
from the finite rotations head model with those from a recently presented finite element head model \cite{Carlsen2021}. In \S\ref{sec:smallrotation}, we compare the  finite rotations head model with the surveyed ICM head models.
We make a few concluding remarks in \S\ref{sec:con}.

\section{Mechanics and mathematical preliminaries}
\label{sec:preliminaries}
In this section we briefly present the mechanical and mathematical preliminaries from \cite[\S2.1]{wan2022, Rahaman2020} that are needed for the development of the proposed head model.
\subsection{Notation}
We denote the space of real numbers as $\mathbb{R}$,  the set of natural numbers as $\mathbb{N}$,
the set of non-negative real numbers as $\mathbb{R}_{\geq 0}$, the set of positive real numbers as $\mathbb{R}_{> 0}$, and the set of non-negative integers as $\mathbb{Z}_{\geq 0}$.

An n-dimensional multi-index is an n-tuple, $n\in\mathbb{N}$, defined as
\begin{equation}
\alpha=\pr{\alpha_1,\alpha_2,\cdots,\alpha_n},
\end{equation}
where $\alpha_i \in \mathbb{Z}_{\geq 0}$, $i\in \pr{1,2,\cdots,n}$. The partial derivative of a function $\pr{x_1,x_2,\cdots,x_n} \mapsto f\ag{x_1,x_2,\cdots,x_n}$ w.r.t. $\alpha$ is defined as
\begin{equation}
\partial ^{\alpha}f=\partial^{\alpha_1}_1\partial^{\alpha_2}_2\cdots\partial^{\alpha_n}_n f,
\end{equation}
where $\partial^{\alpha_i}_i:=\partial^{\alpha_i}/\partial x_i^{\alpha_i}$.

\subsection{Geometry of the abstract and physical spaces in our model}
\label{sec:GeometryinFRSS2DElasticModel}
Let $\mathbb{E}_{\rm R}$ be a finite dimensional, oriented, Hilbert space,
i.e., a Euclidean vector space, and let the affine point space $\mathcal{E}_{\rm R}$ have $\mathbb{E}_{\rm R}$ as its associated vector translation space.  We refer to $\mathbb{E}_{\rm R}$ and $\mathcal{E}_{\rm R}$ as the reference Euclidean vector and point space, respectively.
Let $\mathbb{E}$ and $\mathcal{E}$ be another pair of Euclidean vector and affine point space, respectively.
The topological space $\mathcal{B}$ serves as our model for the brain that executes its motion in $\mathcal{E}$. For that reason, we refer to $\mathbb{E}$ and $\mathcal{E}$ as the physical Euclidean vector space and point space, respectively.

\begin{figure}[t]
    \centering
        \includegraphics[width=\textwidth]{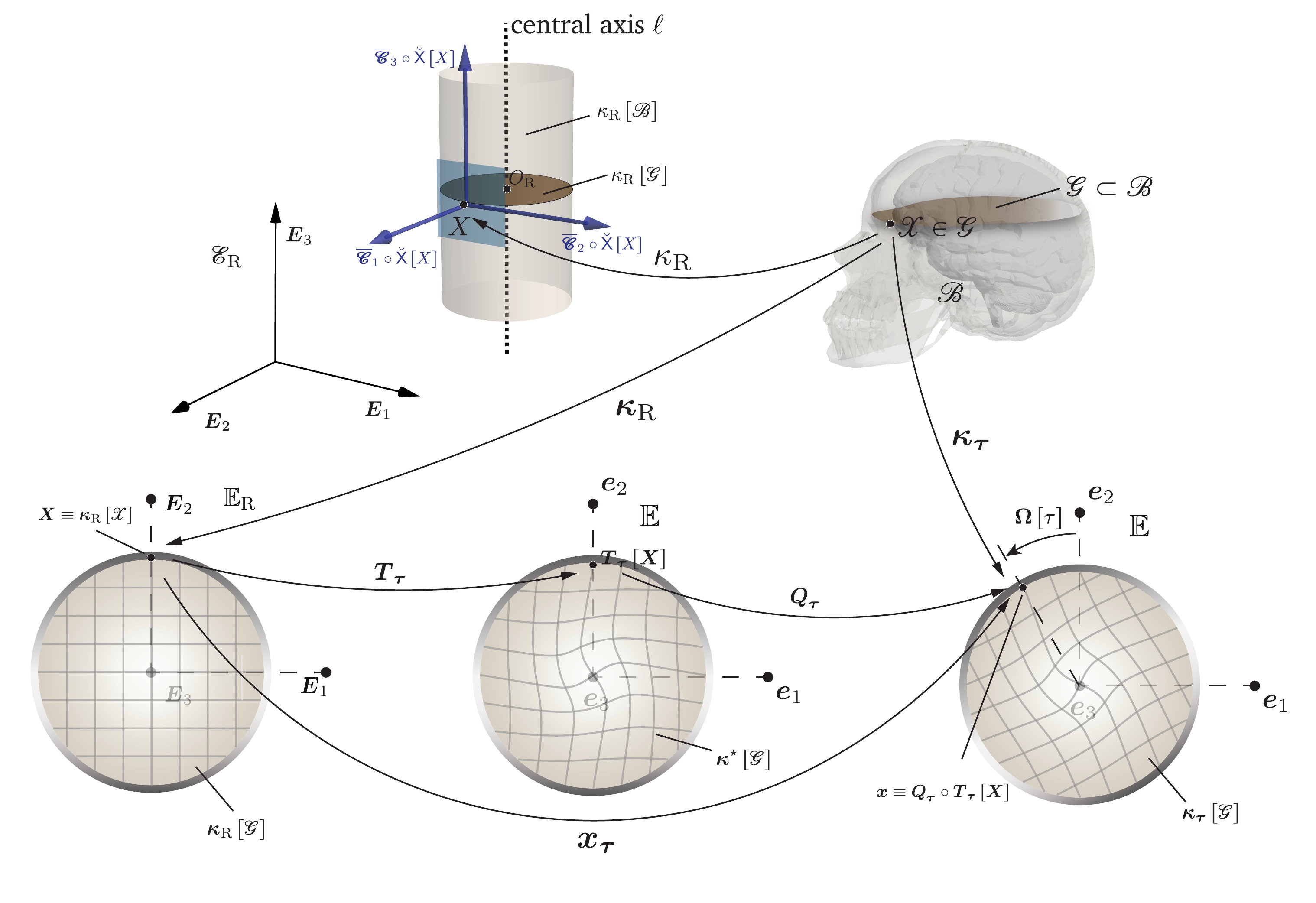}
    \caption{
    An illustration of the various mechanics and mathematical objects that we use in the construction of our finite rotations head model. All the mathematical symbols in this illustration are defined in \S\ref{sec:GeometryinFRSS2DElasticModel}, \S\ref{sec:KinematicsinFRSS2DElasticModel}, and \S\ref{sec:MLPS95Def}.
  }
    \label{fig:notion}
\end{figure}

We call a select continuous, injective map from $\mathcal{B}$ into $\mathbb{E}_{\rm{R}}$ the reference configuration and denoted it as $\boldsymbol{\kappa}_{\rm R}$.
The elements of $\mathcal{B}$ are called material particles.
We call $\boldsymbol{X}\equiv\boldsymbol{\kappa}_{\rm R}\ag{\mathcal{X}}$ the particle $\mathcal{X}$'s reference position vector and $\boldsymbol{\kappa}_{\rm R}\ag{\mathcal{B}}$ the reference body (see Fig.~\ref{fig:notion}).
Taking some arbitrary point $O_{\rm R}\in\mathcal{E}_{\rm R}$ to be $\mathcal{E}_{\rm R}$'s origin, to $\u{\kappa}_{\rm R}$ we associate the map $\kappa_{\rm R}: \mathcal{B} \rightarrow \mathcal{E}_{\rm R}$ such that $O_{\rm R}+\u{\kappa}_{\rm R}\ag{\mathcal{X}}=\kappa_{\rm R}\ag{\mathcal{X}}$. We call $X\equiv\kappa_{\rm R}\ag{\mathcal{X}}$ the particle $\mathcal{X}$'s reference point.

\paragraph{\textbf{Cartesian basis vectors}}
The sets $\pr{\u{E}_{i}}_{i\in\mathcal{I}}$ and $\pr{\u{e}_{i}}_{i\in\mathcal{I}}$, where $\mathcal{I}:=\pr{1,2,3}$, are orthonormal sets of basis vectors for $\mathbb{E}_{{\rm R}}$ and $\mathbb{E}$, respectively.
By orthonormal we mean that the inner product between $\u{E}_i$ and $\u{E}_j$,
or $\u{e}_{i}$ and $\u{e}_j$, where $i, j\in\mathcal{I}$, equals $\delta_{ij}$,
the Kronecker delta symbol, which equals unity iff $i=j$ and zero otherwise.
In our problem, we take $\u{E}_{i}$ and $\u{e}_{i}$, $i\in\mathcal{I}$, to have the units of meters.
The Cartesian co-ordinates of $X$ which we denote as $\breve{\sf{X}}\ag{X}=\pr{\breve{\sf{X}}_i\ag{X}}_{i\in\mathcal{I}}$, are components of $\u{X}$ w.r.t. $\u{E}_i$, that is $\breve{\sf{X}}_i\ag{X}=X_i$, where $X_i:=\u{X}\cdot\u{E}_i$.
For simplicity, ${\sf X} \equiv \pr{X_1,X_2,X_3}$.
We denote the space of all $m \times n$ real nested ordered sets, where $m, n\in \mathbb{N}$, $\mathcal{M}_{m \times n}(\mathbb{R})$.
Thus $\breve{\sf{X}}\ag{X}\in\mathcal{M}_{3 \times 1}(\mathbb{R})$.
We call the map $\mathcal{E}_{\rm{R}}$ $ \ni X \mapsto \breve{\mathsf{X}}\ag{X}\in \mathcal{M}_{3 \times 1}(\mathbb{R})$ the Cartesian co-ordinate map.
When we refer to $\u{X} \in \mathbb{E}_{\rm R}$, ${\sf X}\in \mathcal{M}_{3 \times 1}(\mathbb{R})$, or $X \in \mathcal{E}_{\rm R}$ as a material particle we in fact mean the
material particle $\mathcal{X}\in\mathcal{B}$.

\paragraph{\textbf{Cylindrical basis vectors}}
The cylindrical co-ordinates of ${\sf X}$, which we denote as $(\bar{r}\ag{\sf X}, \bar{\theta}\ag{\sf X}, \bar{z}\ag{\sf X})$, are defined in the standard manner using the Cartesian co-ordinates~${\sf X}$,
\begin{subequations}
\begin{align}
\bar{r}\ag{{\sf X}} & = \sqrt{X_1^2+X_2^2},\\
\bar{\theta}\ag{\sf X} & = {\sf atan2}\ag{X_2,X_1}, \\
\bar{z} \ag{\sf X} & =  X_3.
\end{align}
\end{subequations}
Then the cylindrical basis vectors $\pr{\overline{\physC}_{i}\ag{\mathsf{X}}}_{i\in \mathcal{I}}$ for $\mathbb{E}_{{\rm R}}$ are defined as
\begin{subequations}
\label{eqs:Ci}
\begin{equation}
\pr{\overline{\physC}_{i}\ag{\mathsf{X}}}_{i\in \mathcal{I}}=\bar{\mathsf{R}}\ag{\mathsf{X}}\pr{\u{E}_{i}}_{i\in\mathcal{I}},
\end{equation}
where
\begin{equation}
\bar{\mathsf{R}}\ag{\mathsf{X}}
=
\frac{1}{\sqrt{X_1^2+X_2^2}}\left(\begin{array}{ccc}
X_1 & X_2 & 0 \\ [3 pt]
 -X_2 & X_1 & 0 \\ [3 pt]
 0 & 0 & \sqrt{X_1^2+X_2^2}
\end{array}
\right).
\label{eq:RDef}
\end{equation}
\end{subequations}


Say $\mathbb{W}$ and $\mathbb{U}$ are two arbitrary, oriented, finite dimensional Hilbert spaces; for instance, they can
be $\mathbb{E}_{\rm R}$ and $\mathbb{E}$. We denote the space of all
linear maps (transformations/operators) from $\mathbb{W}$ to $\mathbb{U}$
as $\ensuremath{\mathcal{L}}(\mathbb{W},\mathbb{U})$.
We denote the norm of a vector $\u{w}_{1}$ in $\mathbb{W}$ that
is induced by $\mathbb{W}$'s inner product, i.e., $(\u{w}_{1}\cdot_{\mathbb{W}}\u{w}_{1})^{1/2}$, as $\norm{\u{w}_{1}}_{\mathbb{W}}$. For $\u{u}_{1}\in\mathbb{U}$, the expression $\u{u}_{1}\otimes\u{w}_{1}$
denotes the linear map from $\mathbb{W}$ to $\mathbb{U}$
defined as
\begin{equation}
\pr{\u{u}_{1}\otimes\u{w}_{1}}\u{w}_{2}=\u{u}_{1}\pr{\u{w}_{1}\cdot_{\mathbb{W}}\u{w}_{2}},
\end{equation}
where $\u{w}_{2}\in\mathbb{W}$. If the sets $\pr{\u{u}_i}_{i\in\mathcal{I}}$
and $\pr{\u{w}_i}_{i\in\mathcal{I}}$ provide bases for $\mathbb{U}$
and $\mathbb{W}$, respectively, then it can be shown that $\pr{\pr{\u{u}_{i}\otimes\u{w}_j}_{j\in\mathcal{I}}}_{i\in\mathcal{I}}$,
which we will henceforth abbreviate as $\pr{\u{u}_{i}\otimes\u{w}_j}_{i,j\in\mathcal{I}}$, provides a basis for $\ensuremath{\mathcal{L}}(\mathbb{W},\mathbb{U})$. The symbol $T_{ij}$, where $i,j\in\mathcal{I}$, is called the component of $\u{T}\in\mathcal{L}(\mathbb{W},\mathbb{U})$
w.r.t. $\u{u}_{i}\otimes\u{w}_j$ iff $T_{ij}=\u{u}_{i}\cdot_{\mathbb{U}}\pr{\u{T}\u{w}_j}$. We call the nested ordered set $\pr{T_{ij}}_{i,j\in\mathcal{I}}$ the non-dimensional form of $\u{T}$ w.r.t. $\pr{\u{u}_{i}\otimes\u{w}_j}_{i,j\in\mathcal{I}}$,
and denote it briefly as $\sf{T}$.
We sometimes access the $i^{\rm th}$, $j^{\rm th}$ component of $\sf{T}$,
where $i,j\in\mathcal{I}$, as $\pr{\sf{T}}_{ij}$ or ${\sf T}_{\cdot i \cdot \!j}$. That is, $\pr{\sf{T}}_{ij}=T_{ij}$.
We denote the norm of the operator $\u{T}$ as $\norm{\u{T}}_{\mathbb{U}\otimes \mathbb{W}}$, which is defined as
\begin{equation}
\norm{\u{T}}_{\mathbb{U}\otimes \mathbb{W}}=\sup_{\u{w}_1 \in \mathbb{W},\u{w}_1 \neq \u{0}} \frac{\norm{\u{T}\u{w}_1}_{\mathbb{U}}}{\norm{\u{w}_1}_{\mathbb{W}}}.
\end{equation}
To make some of the ensuing expressions appear less cumbersome, we will omit the subscripts of the $\cdot$ symbol and the $\norm{\cdot}$ operator. Whether we mean $\norm{\cdot}_{\mathbb{W}}$ or $\norm{\cdot}_{\mathbb{U}\otimes \mathbb{W}}$ will be clear from the argument of $\norm{\cdot}$.


\subsection{Kinematics}
\label{sec:KinematicsinFRSS2DElasticModel}
%
%
Select a continuous injective map $\kappa_{\rm R}:\mathcal{B}\to \mathcal{E}_{\rm R}$ such that $\kappa_{\rm R}\ag{\mathcal{B}}$ is a finite right cylinder of radius of $r_0$ meters, where $r_0\in \mathbb{R}_{>0}$, with its central axis $\ell$ passing through $O_{\rm R}$ and parallel to $\u{E}_3$ direction (see Fig.~\ref{fig:notion}).

We model time as a one-dimensional normed vector space $\mathbb{T}$ and denote a typical element in it as $\u{\tau}=\tau\u{s}$, where $\tau\in\mathbb{R}$ and $\u{s}$ is a fixed vector which has units of seconds. 
We model the body's motion using the one-parameter family of maps $\u{x}_{\u{\tau}}\footnote{In \cite[\S2.1]{wan2022} and \cite[\S2.1]{Rahaman2020} $\u{x}_{\u{\tau}}$ (the subscript is set in bold font)  appears as $\u{x}_{\tau}$ (the subscript  is set in regular font).}:\mathbb{E}_{\rm R}\rightarrow \mathbb{E}$ (see Fig.~\ref{fig:notion}).
We call $\u{x}_{\u{\tau}}$ the deformation map and $\u{x}\equiv\u{x}_{\u{\tau}}\ag{\u{X}}$ the material particle $\u{X}$'s position vector at the time instance $\u{\tau}$. The set $\u{\kappa}_{\u{\tau}}\ag{\mathcal{B}}$=$\set{\u{x}_{\u{\tau}}\ag{\u{X}}\in\mathbb{E}}{\u{X}\in\u{\kappa}_{\rm R}\ag{\mathcal{B}}}$ is called the current body.

\paragraph{\textbf{Deformation mapping}}
It can be shown that for our problem for $\tau \ge 0$,
\begin{equation}
  \u{x}_{\u{\tau}}=\u{Q}_{\u{\tau}}\circ\u{T}_{\u{\tau}},
  \label{equ:deformmap}
\end{equation}
where the map $\u{T}_{\u{\tau}}:\mathbb{E}_{\rm R} \rightarrow \mathbb{E}$ is defined by the equation $\u{T}_{\u{\tau}}\ag{\u{X}}=\u{I}\pr{\u{X}+\u{U}_{\u{\tau}}\ag{\u{X}}}$. Here
$\u{I}:=\sum_{i\in\mathcal{I}}\u{e}_{i}\otimes\u{E}_{i}$ and we call the map $\u{U}_{\u{\tau}}:\mathbb{E}_{\rm R} \rightarrow \mathbb{E}_{\rm R}$ the intermediate displacement field of $\mathcal{B}$ at the time instance $\u{\tau}$.
The set $\u{\kappa}^{\star}\ag{\mathcal{B}}$=$\set{\u{T}_{\u{\tau}}\ag{\u{X}}\in\mathbb{E}}{\u{X}\in\u{\kappa}_{\rm R}\ag{\mathcal{B}}}$ is called the intermediate body.


The operator $\u{Q}_{\u{\tau}}$ is a proper (orientation preserving), linear isometry from $\mathbb{E}$ into $\mathbb{E}$. The operator $\u{Q}_{\u{\tau}}$ can be written as $\sum_{i,j\in\mathcal{I}}Q_{ij}\ag{\tau}\u{e}_{i}\otimes\u{e}_{j}$, where $Q_{ij}\in C^{2}\pr{\mathbb{R}_{\ge 0},\mathbb{R}}$.
We abbreviate $\pr{Q_{ij}\ag{\tau}}_{i,j\in\mathcal{I}}\in\mathcal{M}_{3 \times 3}(\mathbb{R})$  as $\sf{Q}\ag{\tau}$. We call $\sf{Q}\ag{\tau}$ the non-dimensional form of $\u{Q}_{\u{\tau}}$ w.r.t. $\pr{\u{e}_{i}\otimes\u{e}_{j}}_{i,j\in\mathcal{I}}$.
Since $\u{Q}_{\u{\tau}}$ is a proper isometry, it follows that $\sf{Q}\ag{\tau}$,
which we refer to as the rotation matrix, belongs to the special orthogonal
group $SO(3)$. As a consequence of belonging to $SO(3)$ the
matrix $\sf{Q}\ag{\tau}$ satisfies the equations

\begin{subequations}
\begin{align}
{\sf Q}^{\sf T}\ag{\tau}\,\sf{Q}\ag{\tau}&=\sf{I},\label{equ:QQ:1}
\intertext{and}
\sf{Q}\ag{\tau}\,{\sf Q}^{\sf T}\ag{\tau}&=\sf{I},\label{equ:QQ:2}
\end{align}
\label{eq:QQ}
\end{subequations}
where ${\sf Q}^{\sf T}\ag{\tau}$ is the transpose of $\sf{Q}\ag{\tau}$, i.e., ${\sf Q}^{\sf T}\ag{\tau}=\pr{\sf{Q}\ag{\tau}}^{\sf T}$ and ${\sf I}=\pr{\delta_{ij}}_{i,j\in\mathcal{I}} \in \mathcal{M}_{3 \times 3}(\mathbb{R})$.


In our problem, the head rotates about $\u{e}_3$ by the angle $\Omega\ag{\tau}$. It can be shown that
\begin{equation}
\sf Q\ag{\tau}
=
\left(\begin{array}{ccc}
\cos \circ~\Omega\ag{\tau}& -\sin \circ~\Omega\ag{\tau} & 0 \\ [3 pt]
 \sin \circ~\Omega\ag{\tau} & \cos \circ~\Omega\ag{\tau} & 0 \\ [3 pt]
 0 & 0 & 1
\end{array}
\right).
\label{eq:QDef}
\end{equation}
We assume the  function $\Omega:[0,\infty)\to \mathbb{R}$ to be at least once continuously differentiable, and for concreteness  that
\begin{subequations}
\doubleequation[eq:OmegaatZeroisZero,eq:OmegaPrimeatZeroisZero]{
\Omega[0]=0,}{
\Omega'[0]=0.}
\label{eq:OmegaInitialConditions}
\end{subequations}
\paragraph{\textbf{Displacements}}
From \eqref{equ:deformmap}, the deformation mapping of the body for $\tau \ge 0$ is described by
\begin{equation}
\u{x}_{\u{\tau}}\ag{\u{X}}=\ou{Q}_{\u{\tau}}\pr{\u{X}+\u{U}_{\u{\tau}}\ag{\u{X}}},
  \label{eq:deformation}
\end{equation}
where $\ou{Q}_{\u{\tau}}:\mathbb{E}_{\rm R} \rightarrow \mathbb{E}$ is defined by the equation $\ou{Q}_{\u{\tau}}=\u{Q}_{\u{\tau}}\u{I}=\sum_{i,j\in\mathcal{I}}Q_{ij}\ag{\tau}\u{e}_{i}\otimes\u{E}_{j}$. In a non-dimensional form \eqref{eq:deformation} reads as
\begin{equation}
    \mathsf{x}_{\tau}\ag{\mathsf{X}}=\mathsf{Q}\ag{\tau}\pr{
    \mathsf{X}+\mathsf{U}_{\tau}\ag{\mathsf{X}}},
\label{eq:deformationND}
\end{equation}
where $\mathsf{x}_{\tau}\ag{\mathsf{X}}:=\pr{x_{\tau i}\ag{\mathsf{X}}}_{i\in \mathcal{I}}$,
$\mathsf{U}_{\tau}\ag{\sf X}:=\pr{U_{\tau i}\ag{\sf X}}_{i\in\mathcal{I}}$,
\begin{align}
{x}_{\tau i}\ag{\sf X}&={\u{x}}_{\u{\tau}}\ag{\sum_{j\in\mathcal{I}}X_j \boldsymbol{E}_j} \cdot \u{e}_i\\
{U}_{\tau i}\ag{\sf X}&={\u{U}}_{\u{\tau}}\ag{\sum_{j\in\mathcal{I}}X_j \boldsymbol{E}_j} \cdot \u{E}_i.
\label{eq:Utaui}
\end{align}

\begin{subequations}
The displacement field is the map $\u{u}_{\u{\tau}}\ag{\cdot}:\mathbb{E}_{\rm R} \rightarrow \mathbb{E}$,
\begin{equation}
\u{u}_{\u{\tau}}\ag{\u{X}}=\u{x}_{\u{\tau}}\ag{\u{X}}-\u{I}\u{X}.
\label{eq:disp}
\end{equation}
In a non-dimensional form \eqref{eq:disp} reads
\begin{equation}
\mathsf{u}_{\tau}\ag{\mathsf{X}}=\mathsf{x}_{\tau}\ag{\mathsf{X}}-\mathsf{X},
\label{eq:dispND}
\end{equation}
where $\mathsf{u}_{\tau}\ag{\mathsf{X}}:=\pr{u_{\tau i}\ag{\mathsf{X}}}_{i\in \mathcal{I}}$ and
\begin{align}
{u}_{\tau i}\ag{\sf X}:={\u{u}}_{\u{\tau}}\ag{\sum_{j\in\mathcal{I}}X_j \boldsymbol{E}_j} \cdot \u{e}_i.
\label{eq:utaui}
\end{align}
\end{subequations}

\paragraph{\textbf{Velocities}}
We call $\mathcal{L}\pr{\mathbb{T},\mathbb{E}}$ the physical velocity
vector space and denote it as $\mathbb{V}$.
It can be shown that the
set $\pr{\u{v}_{i}}_{i\in\mathcal{I}}$, where $\u{v}_{i}\in\mathbb{V}$
and are defined such that $\u{v}_{i}\u{\tau}=\tau\u{e}_i$, that is $\u{v}_{i}:=\u{e}_i\otimes\u{s}^{*}$, where $\u{s}^{*}$ is the dual of $\boldsymbol{s}$, provides
an orthonormal basis for $\mathbb{V}$. The velocity of a material
particle $\u{X}$ executing its motion in $\mathbb{E}$ lies in $\mathbb{V}$.
The velocity of the material particle $\u{X}$ at the instant $\u{\tau}$, which we denote as $\u{V}_{\u{\tau}}\ag{\u{X}}$, equals the value of the Fr\'echet derivative of the map $\mathbb{T}\ni\u{\tau}\mapsto\u{x}_{\u{X}}\ag{\u{\tau}}\in\mathbb{E}$,
where $\u{x}_{\u{X}}\ag{\u{\tau}}=\u{x}_{\u{\tau}}\ag{\u{X}}$, at the time instance $\u{\tau}$.
Thus, it follows from \eqref{eq:deformation} that for $\tau \ge 0$
\begin{equation}
\u{V}_{\u{\tau}}\ag{\u{X}}=\sum_{i,j\in\mathcal{I}}\pr{Q'_{ij}\ag{\tau}X_j+Q'_{ij}\ag{\tau}U_{Xj}\ag{\tau}+Q_{ij}\ag{\tau}U'_{Xj}\ag{\tau}}\u{v}_i.
\label{equ:vel}
\end{equation}
The functions $U_{Xj}:\mathbb{R}_{ \ge 0} \rightarrow \mathbb{R} $ in \eqref{equ:vel} are defined as $U_{Xj}\ag{\tau}=\u{U}_{\u{X}}\ag{\u{\tau}}\cdot\u{E}_j$, where $\u{U}_{\u{X}}\ag{\u{\tau}}=\u{U}_{\u{\tau}}\ag{\u{X}}$, and $Q'_{ij}$ and $U'_{Xj}$ are derivatives of $Q_{ij}$ and $U_{Xj}$, respectively.
Defining $V_{\tau i}\ag{{\sf X}}=\u{V}_{\u{\tau}}\ag{\sum_{j\in\mathcal{I}}X_j \u{E}_j}\cdot \u{v}_i$,
and abbreviating
$\pr{V_{\tau i}\ag{\mathsf{X}}}_{i\in\mathcal{I}}\in\mathcal{M}_{3 \times 1}(\mathbb{R})$,
$\pr{Q'_{ij}\ag{\tau}}_{i,j\in\mathcal{I}}\in\mathcal{M}_{3 \times 3}(\mathbb{R})$, $\pr{U_{Xi}\ag{\tau}}_{i\in\mathcal{I}}\in\mathcal{M}_{3 \times 1}(\mathbb{R})$, and $\pr{U'_{Xi}\ag{\tau}}_{i\in\mathcal{I}}\in\mathcal{M}_{3 \times 1}(\mathbb{R})$ as
$\sf V_{\tau}\ag{\sf X}$,
$\sf Q' \ag{\tau}$,
$\sf U_{\sf X}\ag{\tau}$, and
${\sf U'}_{\sf X}\ag{\tau}$, respectively,  \eqref{equ:vel} in a non-dimensional form reads
\begin{equation}
\sf V_{\tau}\ag{\sf X}=
\mathsf{Q}'\ag{\tau} \mathsf{X}+
\mathsf{Q}'\ag{\tau}\mathsf{U}_{\mathsf{X}}\ag{\tau}+
\mathsf{Q}\ag{\tau}\mathsf{U}'_{\mathsf{X}}\ag{\tau}.
\label{equ:velND}
\end{equation}

\paragraph{\textbf{Accelerations}}
We call $\mathcal{L}\pr{\mathbb{T},\mathbb{V}}$ the physical acceleration
vector space and denote it as $\mathbb{A}$. It can be shown that the
set $\pr{\u{a}_{i}}_{i\in\mathcal{I}}$, where $\u{a}_{i}\in\mathbb{A}$
and are defined such that $\u{a}_{i}\u{\tau}=\tau\u{v}_i$, i.e., $\u{a}_i=\u{v}_i\otimes\u{s}^{*}$, provides
an orthonormal basis for $\mathbb{A}$. The acceleration of a material
particle $\u{X}$ executing its motion in $\mathbb{E}$ lies in $\mathbb{A}$.
The acceleration of $\u{X}$ at the time instance $\u{\tau}$
equals the value of the Fr\'echet derivative of the map $\mathbb{T}\ni\u{\tau}\mapsto\u{V}_{\u{X}}(\u{\tau})\in\mathbb{V}$, where $\u{V}_{\u{X}}(\u{\tau})=\u{V}_{\u{\tau}}(\u{X})$, at the
time instance $\u{\tau}$. Thus, it follows from \eqref{equ:vel}  that for $\tau \ge 0$
\begin{equation}
\u{A}_{\u{\tau}}\ag{\u{X}}=\sum_{i,j\in\mathcal{I}}\pr{Q''_{ij}\ag{\tau}X_j+Q''_{ij}\ag{\tau}U_{Xj}\ag{\tau}+2Q'_{ij}\ag{\tau}U'_{Xj}\ag{\tau}+Q_{ij}\ag{\tau}U''_{Xj}\ag{\tau}}\u{a}_i,
\label{equ:acce}
\end{equation}
where $Q''_{ij}$ and $U''_{Xj}$ are derivatives of $Q'_{ij}$ and $U'_{Xj}$, respectively.
Defining $A_{\tau i}\ag{{\sf X}}=\u{A}_{\u{\tau}}\ag{\sum_{j\in\mathcal{I}}X_j \u{E}_j}\cdot \u{a}_i$, and abbreviating
$\pr{A_{\tau i}\ag{\mathsf{X}}}_{i\in\mathcal{I}}\in\mathcal{M}_{3 \times 1}(\mathbb{R})$, $\pr{Q''_{ij}\ag{\tau}}_{i,j\in\mathcal{I}}\in\mathcal{M}_{3 \times 3}(\mathbb{R})$,  and $\pr{U''_{Xi}\ag{\tau}}_{i\in\mathcal{I}}\in\mathcal{M}_{3 \times 1}(\mathbb{R})$ as  $\sf A_{\tau}\ag{\sf X}$, $\sf Q''\ag{\tau}$, and ${\sf U''}_{\sf X}\ag{\tau}$, respectively,
\eqref{equ:acce} can be written in a non-dimensional form as
\begin{equation}
\sf A_{\tau}\ag{\sf X}=\sf{Q''}\ag{\tau}\sf{X}+\sf{Q''}\ag{\tau}\sf{U}_{\sf X}\ag{\tau}+2\sf{Q'}\ag{\tau}\sf{U'}_{\sf X}\ag{\tau}+\sf{Q}\ag{\tau}\sf{U''}_{\sf X}\ag{\tau}.
\label{equ:acce1}
\end{equation}

\section{Analytical model for head motion and brain deformation}
\label{sec:governequ}

\subsection{Problem statement}
\label{sec:ProblemStatement}
\textit{(A.i)} We model the brain as a \textit{(a)} homogeneous, \textit{(b)} isotropic, \textit{(c)} incompressible, \textit{(d)} elastic solid, and \textit{(e)} the skull as a rigid solid.
\textit{(A.ii)} We will be carrying out a 2D analysis, by which we mean that the components of the displacements in the directions of the rotation axis vanish, the displacements do not vary in the direction of the rotation axis, and the rotation axis remains fixed in space with time.
\textit{(A.iii)} The displacements and deformations in the brain with respect to the skull are taken to be small.
\textit{(A.iv)} The
 displacements of the brain w.r.t. the skull are taken to be axisymmetric.
 The meaning of the assumptions \textit{A.iii} and \textit{A.iv} will be made precise in \S\ref{sec:governingequ}.
\textit{(A.v)} With regard to geometry, we model the brain's cross-sections perpendicular to the rotation axis as disks.
\textit{(A.vi)} We model the interaction between the brain and the skull by positing that the brain's outer surface is rigidly bonded to the skull's interior surface\footnote{Although \emph{in vivo} experiments show that the brain is not rigidly connected to the skull, e.g., in \cite{Feng2010}, this is a common assumption made in ICM head models to simplify the problem.}.
\textit{(A.vii)} The brain and the skull are assumed to be initially at rest or moving with a constant velocity.
\textit{(A.viii)}  Initially the brain is assumed to  not have any displacements with respect to the skull, and
\textit{(A.ix)} to be stress free.

\subsection{Governing equations}
\label{sec:governingequ}

\subsubsection{Displacement-deformation relationship}
It follows from our assumptions \textit{(A.ii)} (2D deformations) and \textit{(A.iv)} (axi-symmetric deformations) in \S\ref{sec:ProblemStatement} that there exist maps $U^{\mathcal{C}}_i$, $i\in \mathcal{I}$, from $\ag{0,r_0}\times\left[0,\infty \right)$ to $\mathbb{R}$  such that
\begin{equation}
\u{U}_{\u{\tau}}\ag{\sum_{i\in\mathcal{I}}X_i \boldsymbol{E}_i}=
U^{\mathcal{C}}_1\ag{\bar{r}\ag{\mathsf{X}},\tau}\overline{\physC}_1\ag{\mathsf{X}}+
U^{\mathcal{C}}_2\ag{\bar{r}\ag{\mathsf{X}},\tau}\overline{\physC}_2\ag{\mathsf{X}}+
U^{\mathcal{C}}_3\ag{\bar{r}\ag{\mathsf{X}},\tau}\overline{\physC}_3\ag{\mathsf{X}}.
\label{equ:disU1}
\end{equation}

We denote the non-dimensional form of the  deformation gradient of the map that transforms $\boldsymbol{\kappa}_{\rm R}\ag{\mathcal{B}}$ to $\boldsymbol{\kappa}^{\star}\ag{\mathcal{B}}$
as $\mathsf{F}^{\star}[\mathsf{X},\tau]$, or $\mathsf{F}^{\star}_{\tau}[\mathsf{X}]$ for short.
And the non-dimensional form of the deformation gradient of the map that transforms $\boldsymbol{\kappa}_{\rm R}\ag{\mathcal{B}}$ to $\boldsymbol{\kappa}_{\boldsymbol{\tau}}\ag{\mathcal{B}}$ as $\mathsf{F}[\mathsf{X},\tau]$, or $\mathsf{F}_{\tau}[\mathsf{X}]$ for short.
It can be shown that
\begin{subequations}
\begin{align}
\mathsf{F}^{\star}[\mathsf{X},\tau]&=\mathsf{I}+\mathsf{H}^{\star}_{\tau}[\mathsf{X}],
\label{eq:deformgradient}
\\
\mathsf{F}[\mathsf{X},\tau]&=\mathsf{Q}[\tau]\mathsf{F}^{\star}[\mathsf{X},\tau].
\label{eq:deformgradient1}
\end{align}
\label{eq:DeformationGradient}
\end{subequations}
In \eqref{eq:deformgradient}, ${\sf H}^{\star}_{\tau}\ag{\sf X}:={\sf Grad}_{\sf X}\ag{{\sf U}_{\tau}}$; here ${\sf Grad}_{\sf X}\ag{{\sf U}_{\tau}}$ is the gradient of ${\sf U}_{\tau}$ at the material particle ${\sf X}$. More explicitly,
\begin{equation}
    \mathsf{H}^{\star}_{\tau}[\mathsf{X}]_{\cdot i\cdot j}=\frac{\partial U_{\tau i}}{\partial X_{j}}\ag{\mathsf{X}}.
\label{eq:StrainDisplacement}
\end{equation}
In \S\ref{sec:ProblemStatement}, we stated that we assume the displacements and deformations of the brain with respect to the skull are small, by which we mean that we assume $\mathsf{U}_{\tau}\ag{\mathsf{X}}$ and $\mathsf{H}^{\star}_{\tau}\ag{\mathsf{X}}$ to be small.


It follows from  our assumption \textit{A.ii} (2D analysis, see \S\ref{sec:ProblemStatement}) and \eqref{eq:deformation} that
\begin{equation}
U_{\tau 3}\ag{\mathsf{X}}=U^{\mathcal{C}}_3\ag{\bar{r}\ag{\mathsf{X}},\tau}=0.
\label{eq:U3e=0}
\end{equation}

Our assumptions that $\mathcal{B}$ is incompressible (assumption \textit{A.i.c} in \S\ref{sec:ProblemStatement}), and $\mathsf{H}^{\star}_{\tau}\ag{\mathsf{X}}$ is small; equation  \ref{eq:U3e=0}; and our assumption of axisymmetric deformation (assumption \textit{A.iv} in \S\ref{sec:ProblemStatement}), i.e., that $\u{U}_{\u{\tau}}\ag{\boldsymbol{X}}$  has the form given in \eqref{equ:disU1} lead us to the result that $U^{\mathcal{C}}_1=0$ (see \S\ref{sec:incom} for details). It then follows from \eqref{equ:disU1} and \eqref{eq:U3e=0} that

\begin{align}
\u{U}_{\u{\tau}}\ag{\sum_{i\in\mathcal{I}}X_i \boldsymbol{E}_i}&=
U^{\mathcal{C}}_2\ag{\bar{r}\ag{\mathsf{X}},\tau}\overline{\physC}_2\ag{\mathsf{X}},
\label{eq:disU3D}
\end{align}
or equivalently that
\begin{align}
\mathsf{U}_{\tau}\ag{\mathsf{X}}&=U^{\mathcal{C}}_2\ag{\bar{r}\ag{\mathsf{X}},\tau}\mathsf{c}_2\ag{\mathsf{X}}.
 \label{eq:disU3ND}
\end{align}
The vector  $\mathsf{c}_2\ag{\mathsf{X}}$ appearing in \eqref{eq:disU3ND} is defined through the equations
\begin{align}
\mathsf{c}_i\ag{\mathsf{X}}&=\pr{\overline{\physC}_{i}\ag{\mathsf{X}}\cdot \boldsymbol{E}_j}_{j\in \mathcal{I}},
\label{eq:Defc2}
\end{align}
$i\in \mathcal{I}$.

\subsubsection{Equation of motion}
Let $\mathsf{P}^{\star}[\mathsf{X},\tau]$ and $\mathsf{P}[\mathsf{X},\tau]$ be the non-dimensional form of $1^{\rm st}$ Piola-Kirchhoff stress tensors in the body when it, respectively, assumes the configurations $\boldsymbol{\kappa}^{\star}$ and  $\boldsymbol{\kappa}_{\rm \tau}$. It follows from the \textit{principle of material frame indifference} (e.g., see \cite[\S21]{Gurtin1982}) that
\begin{equation}
\mathsf{P}[\mathsf{X}, \mathsf{\tau}]=\mathsf{Q}[\tau]\mathsf{P}^{\star}[\mathsf{X}, \mathsf{\tau}].
\label{eq:MFI}
\end{equation}

From Hamilton's principle \cite{Lew2003} and \eqref{eq:MFI} we get $\mathcal{B}$'s equation of motion to be

\begin{equation}
\mathsf{Q}^{\sf T}\ag{\tau}\pr{\sf {Div}_{\mathsf{X}}\ag{\sf{Q}[\tau]\mathsf{P}^{\star}[\tau]}-\rho_{0} \sf{A}_{\tau}\ag{\sf X}}\cdot  \mathsf{c}_{2}\ag{\mathsf{X}} =0,
\label{equ:governing}
\end{equation}
where $\sf {Div}$ is the divergence operator, and $\rho_{0}\in \mathbb{R}_{>0}$ is defined such that $\rho_0~\rm{kg/m^3}$ is the density of $\mathcal{B}$ in the reference configuration\footnote{We do not provide the details of the derivation of \eqref{equ:governing} from Hamilton's principle. The primary novelty in our derivation is that in it the space of admissible displacements and admissible variations are different from those in standard application of Hamilton's principle in finite elasticity as a consequence of our displacements having to satisfy the constraint \eqref{eq:disU3ND}.}.

Let $\bar{\mathsf{P}}_{\mathsf{X}}: \mathcal{M}_{ 3 \times 3}(\mathbb{R}) \to \mathcal{M}_{ 3 \times 3}(\mathbb{R})$ be the material's constitutive equation,
that is $\mathsf{P}^{\star}[\mathsf{X},\mathsf{\tau}]=\bar{\mathsf{P}}_{\mathsf{X}}\ag{\mathsf{F}^{\star}\ag{\mathsf{X},\tau}}$.
We assume that the reference configuration $\boldsymbol{\kappa}_{\rm R}\ag{\mathcal{\mathcal{B}}}$ is stress free (assumption \textit{A.ix} in \S\ref{sec:ProblemStatement}).
It then follows that as $\mathsf{H}\in \mathcal{M}_{3\times 3}\pr{\mathbb{R}}\to \mathsf{o}$, the zero element in $\mathcal{M}_{3\times 3}\pr{\mathbb{R}}$,
\begin{align}
\bar{\mathsf{P}}_{\mathsf{X}}\ag{\mathsf{I}+\mathsf{H}}=\mathsf{C}_{\mathsf{X}}\mathsf{H}+o\pr{\mathsf{H}},
\label{eq:linearziation}
\end{align}
where $\mathsf{C}_{\mathsf{X}}\in \mathcal{M}_{3 \times 3 \times 3 \times 3}(\mathbb{R})$ is a non-dimensional form of the \textit{elasticity tensor}. In \eqref{eq:linearziation} $``o"$ is the Landau ``little-o" symbol that implies that \eqref{eq:linearziation} is equivalent to
\begin{equation}
\lim_{\mathsf{H}\to \mathsf{o}}
    \frac{\lVert \bar{\mathsf{P}}_{\mathsf{X}}\ag{\mathsf{I}+\mathsf{H}}-\mathsf{C}_{\mathsf{X}}\mathsf{H}\rVert}{\norm{\mathsf{H}}}=0.
\label{eq:linearziation2}
\end{equation}
Thus, the linear map  $\mathsf{C}_{\mathsf{X}}$ is the Fr\'echet derivative of $\bar{\mathsf{P}}_{\mathsf{X}}$ at $\mathsf{I}$. In this paper we model the brain as a homogenous solid. This implies that the elasticity tensor does not depend on  $\mathsf{X}$. Therefore, from here on we denote $\mathsf{C}_{\mathsf{X}}$ simply as $\mathsf{C}$.

Assuming $\mathsf{H}^{\star}_{\tau}$ to be uniformly small,
i.e, in the limit of the  deformations of the brain with respect to the skull vanishing uniformly over the brain, we get from \eqref{equ:governing} and \eqref{eq:linearziation2} that
\begin{equation}
\mathsf{Q}^{\sf T}\ag{\tau}\pr{\sf {Div}_{\mathsf{X}}\ag{\sf{Q}[\tau]\mathsf{C} \mathsf{H}^{\star}_{\tau}}-\rho_{0} \sf{A}_{\tau}\ag{\sf X}}\cdot \mathsf{c}_2\ag{\mathsf{X}}=0.
\label{eq:CMeqSimple}
\end{equation}


As we stated in \S\ref{sec:ProblemStatement}, we take $\mathcal{B}$ to be isotropic. For isotropic materials it can be shown that
\begin{subequations}
\begin{align}
\mathsf{C}&=(C_{ijkl})_{i,j,k,l\in \mathcal{I}},
\label{equ:constitutiveNDF}
\\
 C_{ijkl}&=\lambda\delta_{ij}\delta_{kl}+\mu\pr{\delta_{ik}\delta_{jl}+\delta_{il}\delta_{jk}},
 \label{equ:constitutiveCF}
 \end{align}
\label{equ:constitutive}
\end{subequations}
where $\lambda,\ \mu $ are defined such that $\lambda~\rm{N/m^2}$ and $\mu~\rm{N/m^2}$ are $\mathcal{B}$'s Lam\'e parameters.

Writing $\mathsf{C}$ in \eqref{eq:CMeqSimple} in terms of $\lambda$ and $\mu$ using \eqref{equ:constitutive};
$\mathsf{H}^{\star}_{\tau}$ in \eqref{eq:CMeqSimple} in terms of $U^{\mathcal{C}}_2$ using \eqref{eq:StrainDisplacement} and \eqref{eq:disU3ND};
$\mathsf{A}_{\tau}\ag{\sf X}$ in \eqref{eq:CMeqSimple} in terms of
$\mathsf{Q}[\tau]$ and  $\mathsf{U}_{\mathsf{X}}\ag{\tau}$ using \eqref{equ:acce1}; replacing $\mathsf{U}_{\mathsf{X}}\ag{\tau}$ in the resulting equation with $\mathsf{U}_{\tau}\ag{\mathsf{X}}$ and then writing $\mathsf{U}_{\tau}\ag{\mathsf{X}}$ in terms of $U^{\mathcal{C}}_2\ag{\bar{r}\ag{\mathsf{X}},\tau}$  and $\mathsf{c}_2\ag{\mathsf{X}}$ using \eqref{eq:disU3ND};  writing $\mathsf{Q}[\tau]$ in the resulting equation in terms of $\Omega[\tau]$ using \eqref{eq:QDef}; and finally writing $\mathsf{c}_2\ag{\mathsf{X}}$ in terms of $X_1$, $X_2$ in the resulting equation using \eqref{eq:Defc2} and simplifying we get that

\label{eq:ProblemInUC2}
\begin{equation}
  \mu\pr{\partial^{\pr{2,0}} U^{\mathcal{C}}_2\ag{r,\tau}+\frac{\partial^{\pr{1,0}} U^{\mathcal{C}}_2 \ag{r,\tau}}{r}-\frac{U^{\mathcal{C}}_2\ag{r,\tau}}{r^2}}=\rho_0 r \Omega''\ag{\tau}+\rho_0 \partial^{\pr{0,2}} U^{\mathcal{C}}_2\ag{r,\tau}-\rho_0\Omega'\ag{\tau}^2U^{\mathcal{C}}_2\ag{r,\tau}.
  \label{equ:ge21}
\end{equation}

\subsubsection{Boundary and initial conditions}

From \eqref{eq:deformationND}, \eqref{eq:dispND}, and \eqref{eq:disU3ND} it follows that the displacements
\begin{equation}
    \mathsf{u}_{\tau}\ag{\mathsf{X}}
    =\sum_{i=1}^{2}u_i^{\mathcal{C}}\ag{\bar{r}\ag{\mathsf{X}},\tau}\mathsf{c}_i\ag{\mathsf{X}},
    \label{eq:uintermsofCylComponents}
    \end{equation}
where
\begin{subequations}
\begin{align}
u_1^{\mathcal{C}}\ag{r,\tau}&:= r\pr{ \cos\circ~ \Omega\ag{\tau}-1}-\sin\circ~ \Omega\ag{\tau} U_2^{\mathcal{C}}\ag{r,\tau},\\
u_2^{\mathcal{C}}\ag{r,\tau}&:=
    r \sin\circ~ \Omega\ag{\tau}+\cos\circ~ \Omega\ag{\tau} U_2^{\mathcal{C}}\ag{r,\tau}.
\end{align}
\label{eq:DispsCylindericalInTermsofU2C}
\end{subequations}

It follows from our assumptions  \textit{A.viii} (no initial displacements of the brain w.r.t. the skull) that
\begin{subequations}
\begin{equation}
U^{\mathcal{C}}_2\ag{r, \tau=0}=0.
\label{equ:ic1}
\end{equation}
It follows from our assumption that \textit{A.vii} (the brain and skull are initially at rest) and \eqref{eq:uintermsofCylComponents} that at $\tau = 0$,  $\partial^{\pr{0,1}}u_{i}^{\mathcal{C}}\ag{r,\tau}$, where $i=1,2$, vanish.
It therefore follows from \eqref{eq:DispsCylindericalInTermsofU2C} and \eqref{eq:OmegaInitialConditions} that
\begin{equation}
\partial^{\pr{0,1}}   U^{\mathcal{C}}_2\ag{r,\tau=0}=0.
\label{equ:ic2}
\end{equation}
\label{eq:NDUnScaledICs}
\end{subequations}
It follows from our assumptions \textit{A.i.e}  (skull is a rigid solid), and \textit{A.vi} (brain is rigidly connected to the skull) and \eqref{eq:disU3ND} that
\begin{equation}
U^{\mathcal{C}}_2\ag{r=r_0,\tau}=0,
\label{eq:NDUnScaledBCs}
\end{equation}
where recall that $r_0$ is the non-dimensional radius of the brain, which we have modeled as a finite cylinder (see Fig.~\ref{fig:notion}).

\section{Semi-analytical solution}
\label{sec:anasol}

\subsection{Scaling of the equation}
\begin{figure}[t!]
    \centering
        \includegraphics[width=0.95\textwidth]{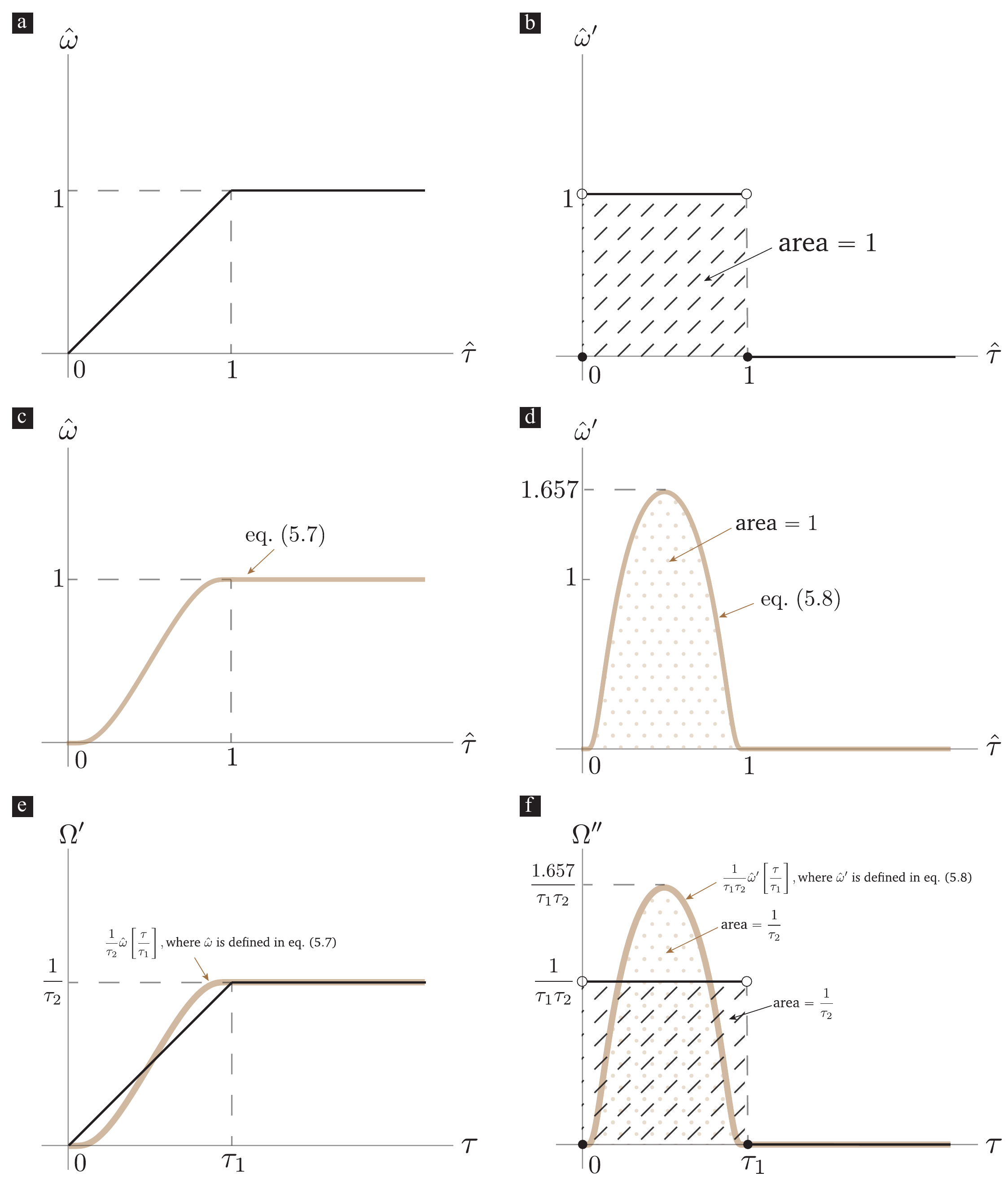}
    \caption{
    The subfigures (a) and (c) show two examples for the scaled loading function $\hat{\omega}\ag{\cdot}$. The scaled loading function is defined in \eqref{equ:scaledw}, in terms of the derivative of the loading function $\Omega\ag{\cdot}$. The function $\Omega\ag{\cdot}$ is defined and discussed in \S\ref{sec:KinematicsinFRSS2DElasticModel}.
    The subfigure (a) shows the graph of the function $[0,\infty)\ni x\mapsto H\ag{x - 1} (1 - x) + x H\ag{x}\in \mathbb{R}$, where $H\ag{\cdot}$ is the unit step function, while subfigure (c) shows the graph of the function \eqref{equ:w2}.
    The subfigure (b) (resp. (d)) shows the graph of the derivative of the function plotted in subfigure (a) (resp. (c)).
     The subfigure (e) shows sketches of the graphs of the  $\Omega'\ag{\cdot}$ that correspond to the scaled loading functions $\hat{\omega}\ag{\cdot}$ plotted in subfigures (a) and (c).
    The curve in black corresponds to the function plotted in subfigure (a), while the  curve in yellow/gold corresponds to the function plotted in subfigure (c).
    The black curve (resp. yellow/golden curve) in subfigure (f) is a sketch of the graph of the derivative of the function plotted in subfigure (e) in black (resp. yellow/gold).
  }
    \label{fig:bump}
\end{figure}

We scale the equations~\eqref{equ:ge21}, \eqref{eq:NDUnScaledICs}, and \eqref{eq:NDUnScaledBCs} by using one spatial constant and three time constants.
The spatial constant is $r_0$, where recall that $r_0$ meters is the radius of our cylindrical head model.
There are three intrinsic time scales in the problem.
Two of them are related to the loading, $\Omega\ag{\cdot}$.
These are $\tau_1$, $\tau_2\in \mathbb{R}_{>0}$, where $\tau_1$ is the non-dimensional time at which $\Omega'\ag{\cdot}$ attains its maximum absolute value, and  $\tau_2$ is the inverse of $\Omega'\ag{\cdot}$'s maximum absolute value.
The time constants $\tau_1$ and $\tau_2$ are illustrated for a model  $\Omega\ag{\cdot}$ in Fig.~\ref{fig:bump}(a). The derivative of that $\Omega\ag{\cdot}$ is sketched in Fig.~\ref{fig:bump}(b).
The last time scale $\tau_s$ seconds is related to the elasticity and dimensions of our head model.
It is defined as
\begin{equation}
  \tau_s=\frac{r_0}{\sqrt{\mu/\rho_0}},
  \label{equ:ts}
\end{equation}
where, recall that, $\mu~\rm N/m^2$  is the brain's shear modulus in our model (cf. \eqref{equ:constitutiveCF}), and $\rho_0~\rm{kg/m^3}$ is the brain's density (cf. \eqref{equ:governing}).

We define the scaled solution $\hat{U}^{\mathcal{C}}_2\ag{\cdot,\cdot}: \ag{0,1}\times\mathbb{R}_{\ge 0} \rightarrow \mathbb{R}$ as
\begin{equation}
\hat{U}^{\mathcal{C}}_2\ag{\hat{r},\hat{\tau}}=U^{\mathcal{C}}_2\ag{\hat{r}r_0,\hat{\tau}\tau_1}/r_0.
\label{equ:scaledu2}
\end{equation}
Similarly the  scaled loading $\hat{\omega}\ag{\cdot}:\mathbb{R}_{\ge 0} \rightarrow \mathbb{R}$ is defined as
\begin{equation}
    \hat{\omega}\ag{\hat{\tau}}=\tau_2 \Omega'\ag{\hat{\tau}\tau_1}.
\label{equ:scaledw}
\end{equation}
The $\hat{\omega}\ag{\cdot}$ (resp. $\hat{\omega}'\ag{\cdot}$) that corresponds to the representative model $\Omega'\ag{\cdot}$ (resp. $\Omega''\ag{\cdot}$) shown with a black line in Fig.~\ref{fig:bump}(a) (resp. (b)) is sketched in Fig.~\ref{fig:bump}(c) (resp. (d)).

In terms of $\hat{U}^{\mathcal{C}}_2\ag{\cdot, \cdot}$, $\hat{r}$, and $\hat{\tau}$ the equation of motion \eqref{equ:ge21} reads
\begin{subequations}
  \begin{equation}
\begin{split}
\partial^{\pr{2,0}} \hat{U}^{\mathcal{C}}_2\ag{\hat{r},\hat{\tau}}+\frac{\partial^{\pr{1,0}} \hat{U}^{\mathcal{C}}_2 \ag{\hat{r},\hat{\tau}}}{\hat{r}}-\frac{\hat{U}^{\mathcal{C}}_2\ag{\hat{r},\hat{\tau}}}{\hat{r}^2}=
\frac{\tau^2_s}{\tau^2_1}\Bigg(\frac{\tau_1}{\tau_2}\hat{r} \hat{\omega}'\ag{\hat{\tau}}+&
\partial^{\pr{0,2}} \hat{U}^{\mathcal{C}}_2\ag{\hat{r},\hat{\tau}}-\\
&\pr{\frac{\tau_1}{\tau_2}}^2\hat{\omega}^2\ag{\hat{\tau}}\hat{U}^{\mathcal{C}}_2\ag{\hat{r},\hat{\tau}}\Bigg);
\end{split}
  \label{equ:scalegoverninge2}
  \end{equation}
the boundary condition \eqref{eq:NDUnScaledBCs} reads
\begin{equation}
\hat{U}^{\mathcal{C}}_2\ag{\hat{r}=1,\hat{\tau}}=0;
\label{equ:scalebc22}
\end{equation}
and the initial conditions \eqref{equ:ic1}, \eqref{equ:ic2}, respectively, read
\doubleequation[equ:scaleic21,equ:scaleic22]{
\hat{U}^{\mathcal{C}}_2\ag{\hat{r}, \hat{\tau}=0}=0,}
{\partial^{\pr{0,1}} \hat{U}^{\mathcal{C}}_2\ag{\hat{r},\hat{\tau}=0}=0.}
\label{equ:scalegoverninge2all}
\end{subequations}

\subsection{Solution using eigenfunction expansions and Sturm-Liouville theory}
\begin{figure}
    \centering
        \includegraphics[height=0.9 \textheight]{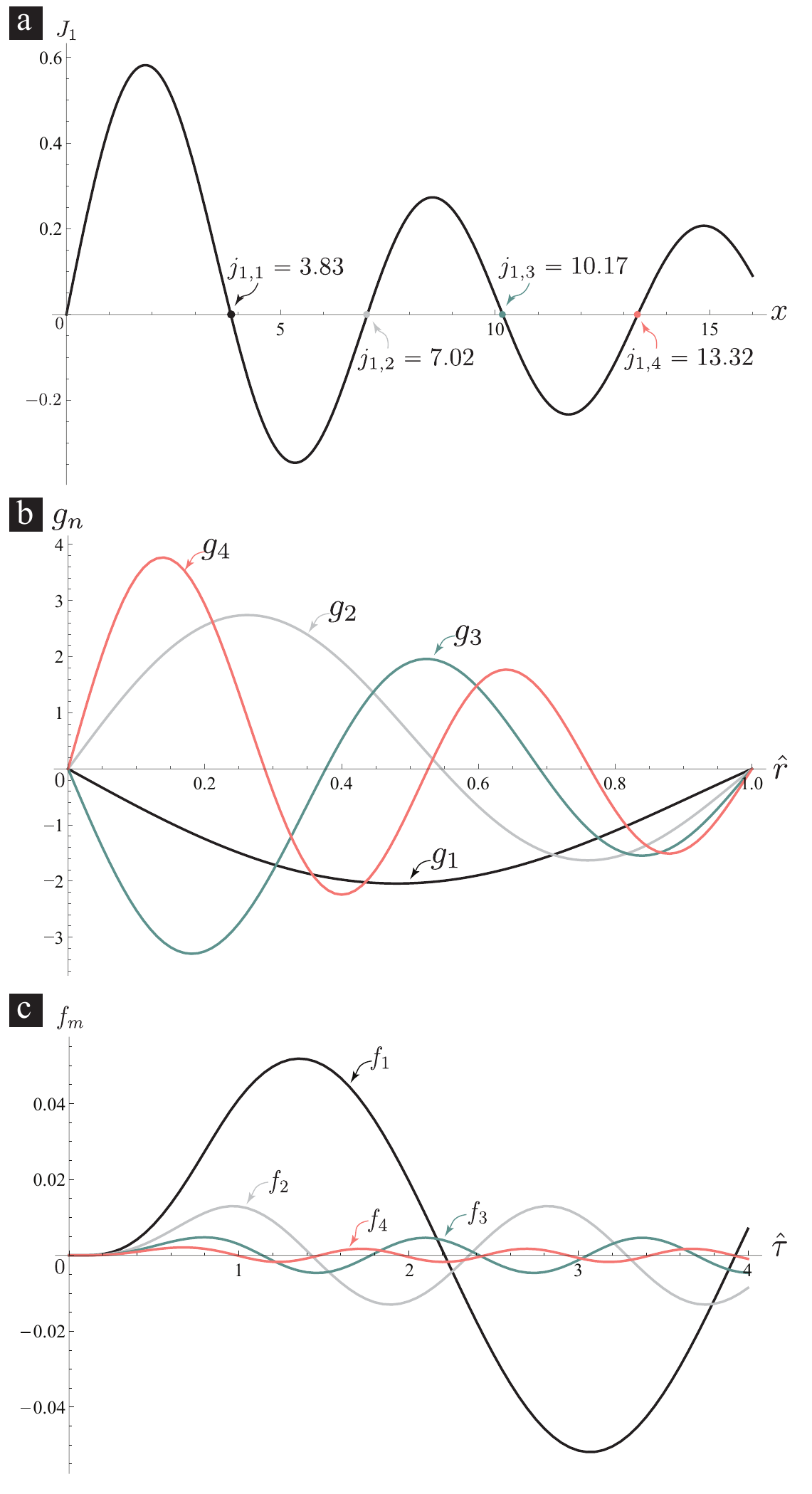}
    \caption{(a) Graph of the Bessel function of the first kind of the first order, $J_1\ag{\cdot}$. The first few zeros of $J_1\ag{\cdot}$, i.e., $j_{1,n}$, $n=1,2,3,4$, are marked and labeled in the plot. (b) Graphs of the first few $g_n\ag{\cdot}$, $n=1,2,3,4$. The functions $g_n\ag{\cdot}$ are defined in \eqref{equ:Bessel}.
    (c) Illustration of the first few $f_m\ag{\cdot}$, $m=1,2,3,4$, that correspond to the material and geometry parameters  \eqref{eq:FEAMatProps}; the loading function $\hat{\omega}\ag{\cdot}$  given in \eqref{equ:w2}; and  $\pr{\tau_1, \tau_2}=\pr{5.5\times10^{-3}, 20\times10^{-3}}$ (equivalently $\pr{\Omega'_{\rm max}, \Omega''_{\rm max}}=\pr{50, 15000}$). These $f_n\ag{\cdot}$ were obtained through a numerical solution of \eqref{equ:timegoverninge2} with the initial conditions \eqref{equ:fncon}.
  }
    \label{fig:fgb}
\end{figure}

In this section we solve for $\hat{U}^{\mathcal{C}}_2\ag{\cdot,\cdot}$ using the method of eigenfunction expansion.
We postulate that $\hat{U}^{\mathcal{C}}_2\ag{\cdot,\cdot}$ can be expressed as,
\begin{equation}
\sum_{n=1}^{\infty} f_n\ag{\hat{\tau}} g_n\ag{\hat{r}},
\label{equ:solution}
\end{equation}
where $f_{n}\ag{\cdot} \in C^{2}(\mathbb{R}_{\ge 0},\mathbb{R})$,
\begin{equation}
 g_n\ag{\hat{r}}:=\sqrt{2}\frac{J_1 \ag{j_{1,n}\hat{r}}}{J_0 \ag{j_{1,n}}}.
 \label{equ:Bessel}
\end{equation}
In equation \eqref{equ:Bessel}, $J_0\ag{\cdot}$ and $J_1\ag{\cdot}$ are Bessel functions of the first kind of zeroth and first order, respectively.
The symbol $j_{1,n}$, $n \in \mathbb{N}$, denotes the n-th zero of $J_1$.
That is, $j_{1,n}$ is defined by the conditions that $J_1\ag{j_{1,n}}=0$ and $j_{1,n}\neq j_{1,m}$ for $m<n$.
The first few $j_{1,n}$ are illustrated in Fig.~\ref{fig:fgb}(a), and the first few $g_n\ag{\cdot}$ are illustrated in Fig.~\ref{fig:fgb}(b).

It can be verified from $g_n\ag{\cdot}$'s definition \eqref{equ:Bessel} that $g_n\ag{1}=0$,  it therefore follows that the form  \eqref{equ:solution} postulated for  $\hat{U}^{\mathcal{C}}_2\ag{\cdot,\cdot}$ satisfies the boundary condition \eqref{equ:scalebc22}.
In order for the form \eqref{equ:solution} to satisfy the initial conditions \eqref{equ:scaleic21} and \eqref{equ:scaleic22} it is necessary and sufficient that
\begin{subequations}
\begin{equation}
f_n\ag{0}=0,
\label{equ:fnc1}
\end{equation}
\begin{equation}
f_n'\ag{0}=0.
\label{equ:fnc2}
\end{equation}
\label{equ:fncon}
\end{subequations}
Next we derive a governing equation for $f_n\ag{\cdot}$, which in conjunction with the initial conditions \eqref{equ:fncon} will determine $f_n\ag{\cdot}$.

Let \begin{equation}
H_r=\left\{f:(0,1)\to \mathbb{R}~| \int_{0}^{1} r f\ag{r}^2\, {\rm d}r<\infty\right\}.
\end{equation}
Defining the inner product   $\pr{\cdot,\cdot}_r:H_r\times H_r \rightarrow \mathbb{R}$,
\begin{equation}
  \pr{f,g}_r=\int_0^1\xi f\ag{\xi}g\ag{\xi}{\rm d}\xi,
  \label{equ:innerproduct}
\end{equation}
on $H_r$, it can be shown that $(H_r, \pr{\cdot, \cdot}_{r})$ is a Hilbert Space. Noting that \begin{equation}
  \pr{g_n,g_m}_r=\delta_{nm},
  \label{equ:gnc2}
\end{equation}
it can be shown that $(g_n\ag{\cdot})_{n\in \mathbb{N}}$ provide an orthonormal basis for $H_r$.

Replacing $\hat{U}^{\mathcal{C}}_2\ag{\hat{r},\hat{\tau}}$ in \eqref{equ:scalegoverninge2} with the form given in \eqref{equ:solution}, replacing the function $(0,1)\ni \hat{r}\mapsto \hat{r}$ that appears in the first term on the right hand side of \eqref{equ:scalegoverninge2} with its expansion in the $g_n\ag{\cdot}$ basis, i.e.,  writing $\hat{r}$ as $\sum_{n\in \mathbb{N}}p_n g_n\ag{\hat{r}}$, where
\begin{equation}
p_n:=\int_0^1\xi^2 g_n\ag{\xi}{\rm d}\xi,
\label{equ:pn}
\end{equation}
in the first term on the right hand side of \eqref{equ:scalegoverninge2}, we get that


\begin{equation}
  \sum_{n=1}^{\infty}f_n\ag{\hat{\tau}}\pr{\mathcal{D}g_n}\ag{\hat{r}}=\sum_{n=1}^{\infty}\frac{\tau_s^2}{\tau_1\tau_2}p_ng_n\ag{\hat{r}}\hat{\omega}'\ag{\hat{\tau}}+\sum_{n=1}^{\infty}\frac{\tau_s^2}{\tau_1^2}f_n''\ag{\hat{\tau}}g_n\ag{\hat{r}}-\sum_{n=1}^{\infty}\frac{\tau_s^2}{\tau_2^2}\hat{\omega}^2\ag{\hat{\tau}}f_n\ag{\hat{\tau}}g_n\ag{\hat{r}},
    \label{equ:scalegoverninge5}
\end{equation}
where $f_n''$ is the derivative of $f_n'$,
and the operator $\mathcal{D}$ is defined such that
\begin{equation}
  \pr{\mathcal{D}g_n}\ag{\hat{r}}=
  g_n''\ag{\hat{r}}+ g_n'\ag{\hat{r}}/\hat{r}-
  g_n\ag{\hat{r}}/\hat{r}^2.
  \label{equ:Doperator}
\end{equation}
Noting that
\begin{equation}
  \pr{\mathcal{D}g_n}\ag{\hat{r}}=-j^2_{1,n} g_n\ag{\hat{r}}
\label{equ:gnc3}
\end{equation}
and substituting $\pr{\mathcal{D}g_n}\ag{\hat{r}}$ in equation \eqref{equ:scalegoverninge5} with the right side of \eqref{equ:gnc3};
taking  inner product (as defined in \eqref{equ:innerproduct}) of the resulting equation with $g_m\ag{\cdot}$, $m\in \mathbb{N}$; and  simplifying the resulting equation using \eqref{equ:gnc2} we get a governing equation of $f_m\ag{\cdot}$ to be
\begin{equation}
  f_m''\ag{\hat{\tau}}+\frac{\tau_1^2}{\tau_s^2}\pr{j_{1,m}^2-\frac{\tau_s^2}{\tau_2^2}\hat{\omega}^2\ag{\hat{\tau}}}f_m\ag{\hat{\tau}}=-\frac{\tau_1}{\tau_2}p_m\hat{\omega}'\ag{\hat{\tau}}.
    \label{equ:timegoverninge2}
\end{equation}

The first few $f_m\ag{\cdot}$ for the $\hat{\omega}\ag{\cdot}$ given in \eqref{equ:w2} are illustrated in Fig.~\ref{fig:fgb}(c). We computed those $f_m\ag{\cdot}$  by numerically integrating \eqref{equ:timegoverninge2} with the initial conditions \eqref{equ:fncon}.
In these numerical calculations   $\tau_1=0.0055$, $\tau_2=0.02$, and $\tau_s=0.0114$. The $p_n$'s, of course are independent of the loading. We computed the $p_n$ by evaluating the integral in \eqref{equ:pn}. The first few  $p_n$ are $p_1=-0.37$, $p_2=-0.20$, $p_3=-0.14$, and $p_4=-0.11$.


\subsection{Displacements, strains, and strain rates}
\label{sec:ThreeTSol}
\begin{figure}
    \centering     \includegraphics[width=0.95\textwidth]{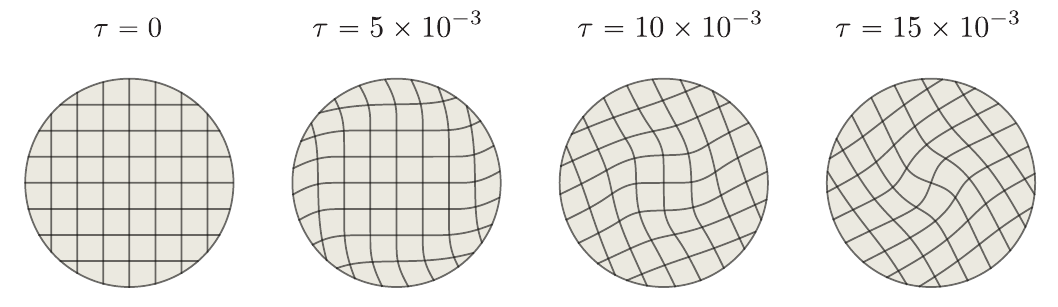}
    \caption{ Configurations of the brain from the
    motion  corresponding to the material and geometry parameters  \eqref{eq:FEAMatProps}; the loading function $\hat{\omega}\ag{\cdot}$  given in \eqref{equ:w2}; and  $\pr{\tau_1, \tau_2}=\pr{5.5\times10^{-3}, 20\times10^{-3}}$ (equivalently $\pr{\Omega'_{\rm max}, \Omega''_{\rm max}}=\pr{50, 15000}$).
    The four configurations shown are for the time instances $\tau=0$, $5\times10^{-3}$, $10\times10^{-3}$, and $15\times10^{-3}$.
    These configurations were computed by  applying the procedure \ref{algo:ThreeTermdisp}.
  }
    \label{fig:reconf}
\end{figure}

\begin{figure}
    \centering     \includegraphics[width=0.95\textwidth]{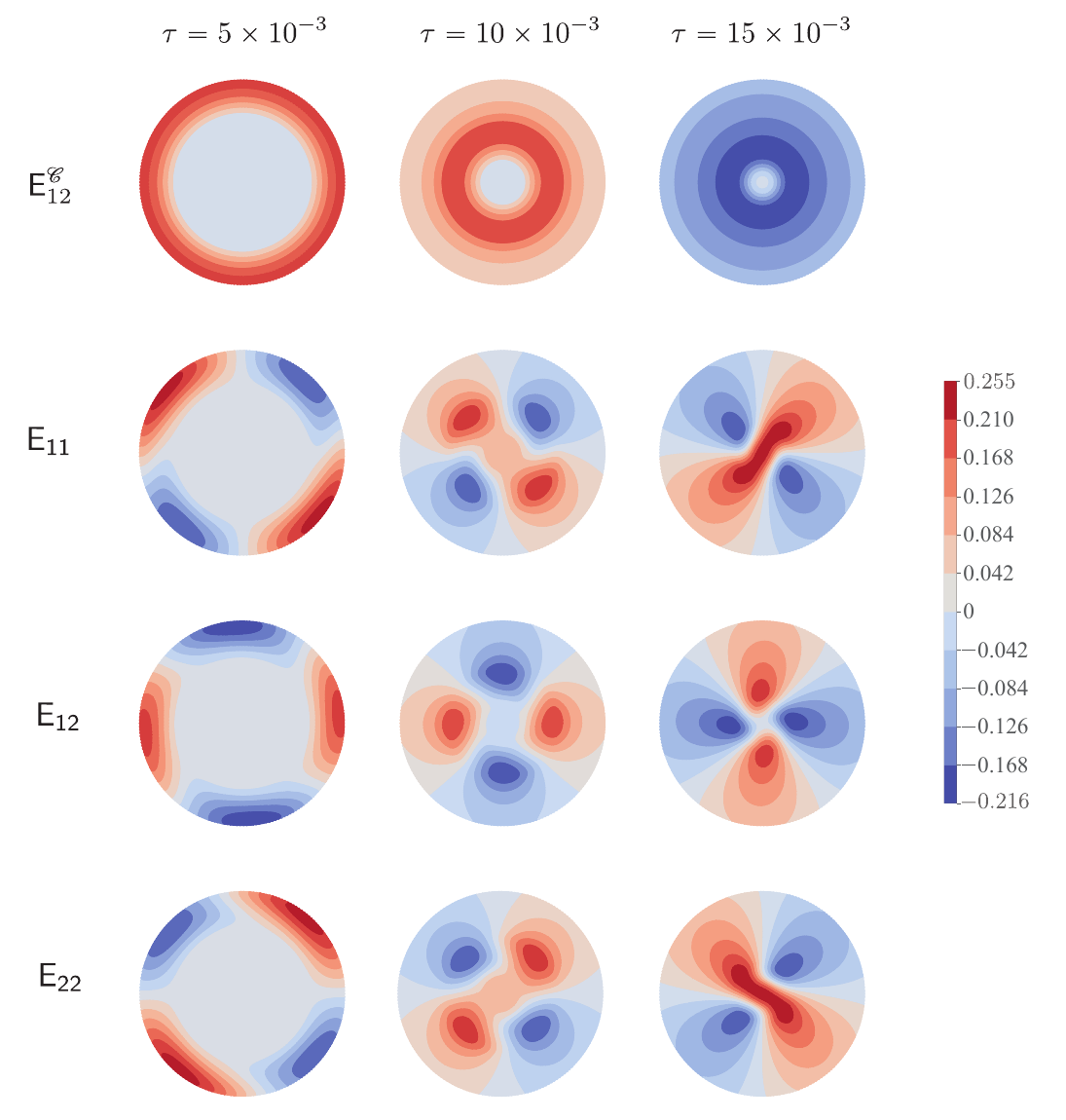}
    \caption{
    Contour plots of the components of the Green-Lagrange strain tensor that corresponds to the motion shown Fig.~\ref{fig:reconf}.
    The first row shows the contour plot of $\mathsf{E}^{\physC}_{12}\ag{\cdot,\cdot}$ (see \eqref{eq:E4}), which is the 1-2 component of the Green-Lagrange strain tensor w.r.t. the   $\pr{\overline{\physC}_{i}\ag{\mathsf{X}}\otimes \overline{\physC}_{j}\ag{\mathsf{X}}}_{i,j\in \mathcal{I}}$ basis.
    The second, third, and fourth rows, respectively, show the contour plots of $\sf E_{11}\ag{\cdot,\cdot}$, $\sf E_{12}\ag{\cdot,\cdot}$, and $\sf E_{22}\ag{\cdot,\cdot}$, which are the Cartesian components of the Green-Lagrange strain tensor (see \eqref{eq:E3}). The first, second, and third columns correspond to the time instances $\tau=5\times10^{-3}$, $10\times10^{-3}$, and $15\times10^{-3}$.
  }
    \label{fig:restrain}
\end{figure}

\begin{figure}
    \centering     \includegraphics[width=0.9\textwidth]{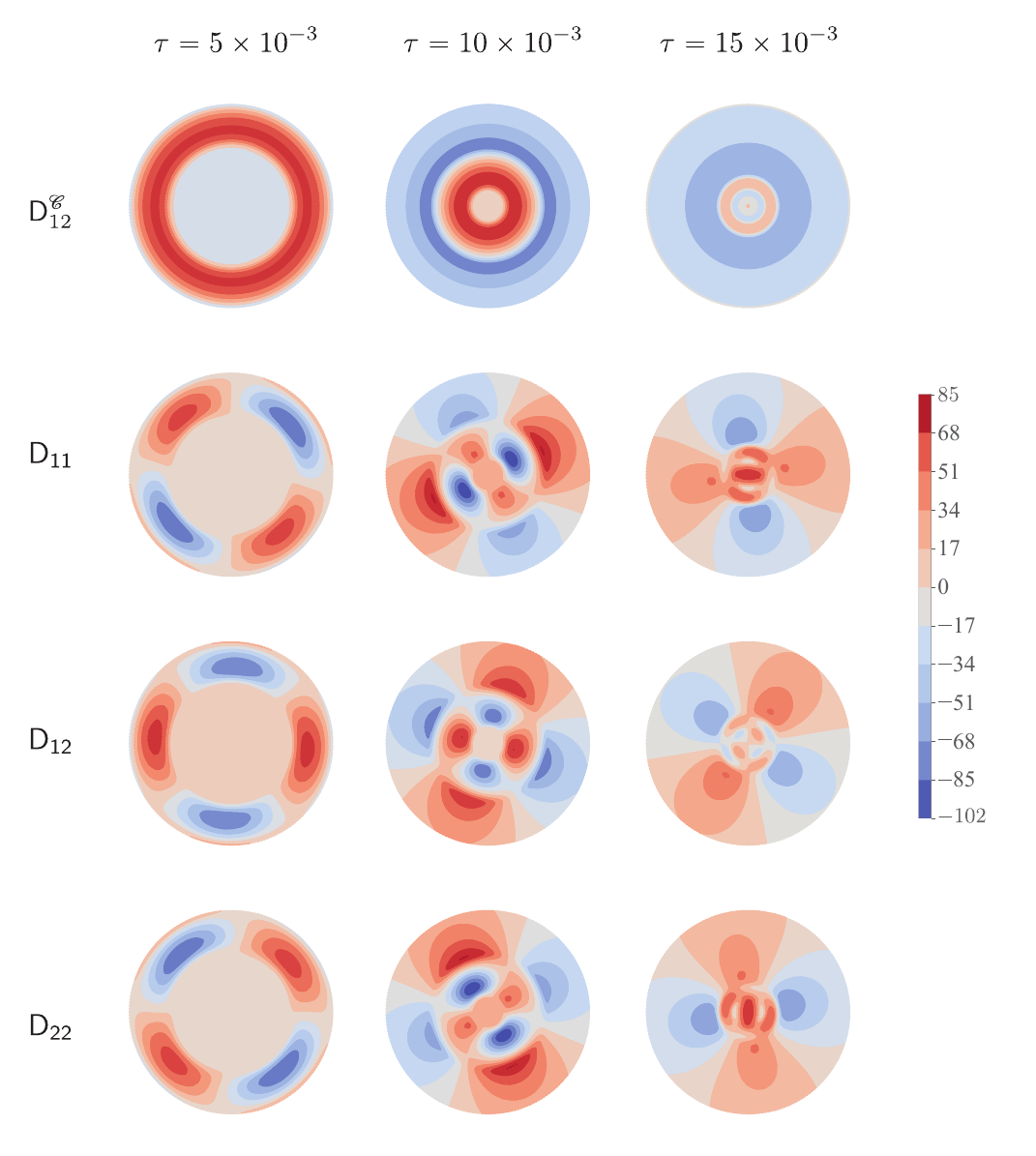}
    \caption{
    Contour plots of the components of the rate of deformation tensor that corresponds to the motion shown in Fig.~\ref{fig:reconf}.
    The first row shows the contour plot of $\mathsf{D}^{\physC}_{12}\ag{\cdot,\cdot}$, which is the 1-2 component of the non-dimensional rate of deformation tensor $\mathsf{D}^{\physC}\ag{\cdot,\cdot}$ (see  \eqref{eq:ER4}).
    The second, third, and fourth rows, respectively, show the contour plots of $\sf D_{11}\ag{\cdot,\cdot}$, $\sf D_{12}\ag{\cdot,\cdot}$, and $\sf D_{22}\ag{\cdot,\cdot}$, which are components of the  non-dimensional rate of deformation tensor $\sf D\ag{\cdot,\cdot}$ (see \eqref{eq:ER1}). The first, second, and third columns correspond to the time instances $\tau=5\times10^{-3}$, $10\times10^{-3}$, and $15\times10^{-3}$.
  }
    \label{fig:restrainrate}
\end{figure}

In this section we outline procedures for computing displacements (\S\ref{sec:disp}), strains (\S\ref{sec:strain}), and strain rates (\S\ref{sec:strainrate}) in our head model. We show the displacements, strains, and strain rates that we computed using these procedures for the representative geometry and material properties given in \eqref{eq:FEAMatProps}, and the loading function $\hat{\omega}\ag{\cdot}$ given in \eqref{equ:w2}, in Figs.~\ref{fig:reconf}, \ref{fig:restrain}, and \ref{fig:restrainrate}, respectively.
 For the geometry and material properties given in \eqref{eq:FEAMatProps} the time scale $\tau_s=11.4 \times 10^{-3}$. We carried out the calculations for the loading related time scales          $\pr{\tau_1, \tau_2}=\pr{5.5\times10^{-3}, 20\times10^{-3}}$ (equivalently $\pr{\Omega'_{\rm max}, \Omega''_{\rm max}}=\pr{50, 15000}$). In each of the  Figs.~\ref{fig:reconf}, \ref{fig:restrain}, and \ref{fig:restrainrate}, we show the calculations for the time instances  $\tau=5 \times 10^{-3}$, $10 \times 10^{-3}$, and $15 \times 10^{-3}$.


\floatname{algorithm}{Procedure}
\subsubsection{Displacements}
\label{sec:disp}
We outline a procedure for computing displacements in our head model in  procedure~\ref{algo:ThreeTermdisp}.
\begin{algorithm}[h]
\caption{Computing displacements in the head model}\label{algo:ThreeTermdisp}
\begin{enumerate}
\item Given an individual's head's biometric and other details, e.g., from MRI scans, select values for the characteristic length scale $r_0$, and the characteristic time scale $\tau_s$.
\item Given a $\Omega\ag{\cdot}$, select the loading times scales $\tau_1$, $\tau_2$ based on the characteristics of $\Omega\ag{\cdot}$.
\item Construct non-dimensional scaled loading functions $\hat{\omega}\ag{\cdot}$ and $\hat{\omega}'\ag{\cdot}$ using the given $\Omega\ag{\cdot}$, the $\tau_1$ and $\tau_2$ from the previous step, and  \eqref{equ:scaledw}.
\item Using \eqref{equ:Bessel} and \eqref{equ:pn} compute $p_n$, $n\in \mathbb{N}$.
\item Using the $\tau_s$, $\tau_1$, $\tau_2$, $\hat{\omega}$, $\hat{\omega}'$, and $p_n$ from the previous steps calculate $f_{m}$, $m\in \mathbb{N}$, by numerically integrating \eqref{equ:timegoverninge2} with the initial conditions given in \eqref{equ:fncon}.
\item Using the calculated $f_m$ and the $g_m$ given by \eqref{equ:Bessel}, construct $\hat{U}^{\mathcal{C}}_2\ag{\cdot, \cdot}$ using \eqref{equ:solution}.
\item Using the $\hat{U}^{\mathcal{C}}_2\ag{\cdot, \cdot}$ from the previous step and \eqref{equ:scaledu2} construct $U^{\mathcal{C}}_2\ag{\cdot, \cdot}$.
\item Compute the (non-dimensional version of the) intermediate displacement field,   $\mathsf{U}_{\tau}$, using  $U^{\mathcal{C}}_2\ag{\cdot, \cdot}$ from the previous step and \eqref{eq:disU3ND}.
\item Compute $\mathsf{x}_{\tau}$ using the $\mathsf{U}_{\tau}$ from the previous step and \eqref{eq:deformationND}, and the displacement of the material particle $\mathsf{X}$ at the time instance $\tau$, as $\mathsf{x}_{\tau}\ag{\mathsf{X}}-\mathsf{X}$.
\end{enumerate}
\end{algorithm}

\subsubsection{Strains}
\label{sec:strain}

For a given loading, head geometry, and material properties, the field  $U^{\mathcal{C}}_2\ag{\cdot, \cdot}$ can be computed by following the first seven steps of procedure \ref{algo:ThreeTermdisp}.
Using that $U^{\mathcal{C}}_2\ag{\cdot, \cdot}$ the intermediate displacement gradients in our head model can be computed as
 \begin{equation}
 \begin{split}
    \mathsf{H}^{\star}_{\tau}\ag{\mathsf{X}}&=\\
    &\left(\begin{array}{ccc}
    \frac{X_1X_2}{\bar{r}^2\ag{\sf X}}\pr{\frac{U^{\physC}_2\ag{\bar{r}\ag{\sf X},\tau}}{\bar{r}\ag{\sf X}}-\partial^{\pr{1,0}} U^{\physC}_2\ag{\bar{r}\ag{\sf X},\tau}}& \frac{-1}{\bar{r}^2\ag{\sf X}}\pr{\frac{X_1^2U^{\physC}_2\ag{\bar{r}\ag{\sf X},\tau}}{\bar{r}\ag{\sf X}}+{X_2^2}\partial^{\pr{1,0}} U^{\physC}_2\ag{\bar{r}\ag{\sf X},\tau}} & 0 \\[4ex]
     \frac{1}{\bar{r}^2\ag{\sf X}}\pr{\frac{X_2^2U^{\physC}_2\ag{\bar{r}\ag{\sf X},\tau}}{\bar{r}\ag{\sf X}}+X_1^2\partial^{\pr{1,0}} U^{\physC}_2\ag{\bar{r}\ag{\sf X},\tau}} & \frac{X_1X_2}{\bar{r}^2\ag{\sf X}}\pr{-\frac{U^{\physC}_2\ag{\bar{r}\ag{\sf X},\tau}}{\bar{r}\ag{\sf X}}+\partial^{\pr{1,0}} U^{\physC}_2\ag{\bar{r}\ag{\sf X},\tau}} & 0\\[4ex]
     0 & 0 & 0
    \end{array}
    \right).
    \label{eq:Hmatrix}
\end{split}
\end{equation}
Various different strain measures can be computed in our head model using the  displacement gradients from \eqref{eq:Hmatrix}.
For example, using \eqref{eq:Hmatrix} it can be shown that the Green-Lagrange strains,
\[
\mathsf{E}\ag{\mathsf{X},\tau}:=\pr{\mathsf{H}^{\star}_{\tau}\ag{\mathsf{X}}+{\mathsf{H}^{\star}_{\tau}}^{\sf T}\ag{\mathsf{X}}+{\mathsf{H}^{\star}_{\tau}}^{\sf T}\ag{\mathsf{X}}\mathsf{H}^{\star}_{\tau}\ag{\mathsf{X}}}/2,
\]
in our head model in terms of $U_2^{\mathcal{C}}\ag{\cdot, \cdot}$ read
\begin{subequations}
\begin{align}
    \mathsf{E}\ag{\mathsf{X},\tau}&=\bar{\sf R}^{\sf T}\ag{\sf X}\mathsf{E}^{\physC}\ag{\bar{r}\ag{\mathsf{X}},\tau}\bar{\sf R}\ag{\sf X}, \label{eq:E3}\intertext{where}
    \mathsf{E}^{\physC}\ag{r,\tau}&:=
  \frac{1}{2}  \left(\begin{array}{ccc}
    \pr{\partial^{\pr{1,0}} U^{\physC}_2\ag{r,\tau}}^2 & \partial^{\pr{1,0}} U^{\physC}_2\ag{r,\tau}-\frac{U^{\physC}_2\ag{r,\tau}}{r}& 0 \\ [3 pt]
    \partial^{\pr{1,0}} U^{\physC}_2\ag{r,\tau}-\frac{U^{\physC}_2\ag{r,\tau}}{r} & \pr{\frac{U^{\physC}_2\ag{r,\tau}}{r}}^2 & 0\\ [3 pt]
     0 & 0 & 0
    \end{array}
    \right).
    \label{eq:E4}
\end{align}
\label{eq:strain}
\end{subequations}
We give details for computing the spatial logarithimic strain tensor in our head model using the $\mathsf{H}^{\star}_{\tau}$ given in \eqref{eq:Hmatrix} in \S\ref{sec:MLPSDef}.

\subsubsection{Strain rates}
\label{sec:strainrate}
Various strain rate measures can be computed using the rate of deformation tensor (or rate of strain tensor).
The rate of deformation tensor is the symmetric part
of the spatial velocity gradient tensor.
Let $\mathsf{D}\ag{\mathsf{X},\tau}$ be the non-dimensional form of the rate of deformation tensor of a material particle $\u{X}$ at time instance $\u{\tau}$ w.r.t. $\pr{\u{e}_i\otimes\u{s}^{*}\otimes\u{e}_j}_{i,j\in \mathcal{I}}$.
After computing the field  $U^{\mathcal{C}}_2\ag{\cdot, \cdot}$ by following the first seven steps of procedure \ref{algo:ThreeTermdisp} we can compute $\mathsf{D}\ag{\mathsf{X},\tau}$   in our head model as
\begin{subequations}
\begin{align}
\mathsf{D}\ag{\mathsf{X},\tau}&=
    {\sf Q}\ag{\tau}\bar{\sf R}^{\sf T}\ag{\sf X}\mathsf{D}^{\physC}\ag{\bar{r}\ag{\mathsf{X}},\tau}\bar{\sf R}\ag{\sf X}{\sf Q}^{\sf T}\ag{\tau},
    \label{eq:ER1}
    \intertext{where}
\mathsf{D}^{\physC}\ag{r,\tau}&:=  \left(\begin{array}{ccc}
      \dfrac{\partial^{\pr{1,0}} U^{\physC}_2\ag{r,\tau}\partial^{\pr{0,1}} U^{\physC}_2\ag{r,\tau}}{r+U^{\physC}_2\ag{r,\tau}\partial^{\pr{1,0}} U^{\physC}_2\ag{r,\tau}}& \dfrac{r \partial^{\pr{1,1}} U^{\physC}_2\ag{r,\tau}-\partial^{\pr{0,1}} U^{\physC}_2\ag{r,\tau}}{2\pr{r+U^{\physC}_2\ag{r,\tau}\partial^{\pr{1,0}} U^{\physC}_2\ag{r,\tau}}}& 0 \\ [4ex]
        \dfrac{r \partial^{\pr{1,1}} U^{\physC}_2\ag{r,\tau}-\partial^{\pr{0,1}} U^{\physC}_2\ag{r,\tau}}{2\pr{r+U^{\physC}_2\ag{r,\tau}\partial^{\pr{1,0}} U^{\physC}_2\ag{r,\tau}}} &   \dfrac{U^{\physC}_2\ag{r,\tau}\partial^{\pr{1,1}} U^{\physC}_2\ag{r,\tau}}{r+U^{\physC}_2\ag{r,\tau}\partial^{\pr{1,0}} U^{\physC}_2\ag{r,\tau}} & 0\\ [4ex]
       0 & 0 & 0
      \end{array}
      \right).
    \label{eq:ER4}
\end{align}
\label{eq:strainrate}
\end{subequations}

\section{Discussion}
\label{sec:dis}



\subsection{Comparison with finite element solutions}
\label{sec:femcompare}

We compared ``maximum 95th percentile MPLS'', which we will define in \S\ref{sec:MLPSDef}--\S\ref{sec:MLPS95Def}, values in our head model with those in a related, finite element analysis (FEA) based, 2D head model \cite{Carlsen2021}. 
The term ``MPLS''  stands for maximum principal logarithmic strain.
The MPLS of a material particle $\u{X}$ at a time instance $\u{\tau}$, which we denoted as $\phi_{\tau1}\ag{\sf X}$, is defined as the maximum eigenvalue of
$\ln\mathsf{V}_{\tau}\ag{\mathsf{X}}$,
the non-dimensional form of (spatial) logarithmic strain tensor at $\mathsf{X}$. 
We recall the definition of $\ln\mathsf{V}_{\tau}\ag{\mathsf{X}}$ and more precisely define $\phi_{\tau1}\ag{\sf X}$ in \S\ref{sec:MLPSDef}. 
In \S\ref{sec:MLPSDef} we also provide the details for computing $\phi_{\tau1}\ag{\sf X}$ in our head model.
We define the  95th percentile MPLS at the time instance $\tau$, which we denote as $\phi^{95}\ag{\tau}$, in \S\ref{sec:MLPS95Def} using $\phi_{\tau1}\ag{\sf X}$.
The maximum 95th percentile MPLS, which we denote as, $\phi^{95}_{\rm max}$, is also defined in \S\ref{sec:MLPS95Def} using $\phi^{95}\ag{\tau}$.

We computed the $\phi^{95}_{\rm max}$ values in our head model for the loading, geometry, and material  properties that are similar to those used in the finite element head model simulations.
We give their details  in  \S\ref{sec:FEM}.
The $\phi^{95}_{\rm max}$ values computed in our model are compared with those in the finite element head model in Fig.~\eqref{fig:95mpsrika}.
As can be seen in Fig.~\eqref{fig:95mpsrika}, the qualitative dependence of $\phi^{95}_{\rm max}$ values on the two loading parameters $\Omega_{\rm max}'$ and $\Omega_{\rm max}''$, which we will detail in \S\ref{sec:FEM}, in our head model is very similar to that seen in the finite element  head model.
As will be detailed in \S\ref{sec:FEM}, the finite element head model is far more detailed and sophisticated than our ICM based head model.
For example, it includes information about the brain's spatially heterogenous and anisotropic material behavior,  as well as information about the brain's viscous and non-linear elastic behavior.
What we find remarkable is that despite the far less sophistication and detail in our model, it is able to predict strains values  that are as close to the finite element ones as they are in Fig.~\eqref{fig:95mpsrika}.

\subsubsection{Maximum principle logarithmic strain (MPLS, $\phi_{\tau 1}\ag{\mathsf{X}}$)}
\label{sec:MLPSDef}
When $\mathsf{F}_{\tau}\ag{\mathsf{X}}$ is invertible it follows from the \textit{polar decomposition theorem} \cite[e.g., see][\S2]{Gurtin1982} that there exists a unique symmetric positive definite tensor $\mathsf{V}_{\tau}\ag{\mathsf{X}}$, called the left stretch tensor, such that $
\mathsf{V}_{\tau}\ag{\mathsf{X}}\mathsf{V}_{\tau}\ag{\mathsf{X}}=\mathsf{B}_{\tau}\ag{\mathsf{X}},
$
where
\begin{equation}
\mathsf{B}_{\tau}\ag{\mathsf{X}}:={\sf F}_{\tau}\ag{\sf X}\pr{\mathsf{F}_{\tau}\ag{\sf X}}^{\sf T}
\label{eq:DefB}
\end{equation} is called the left Cauchy-Green deformation tensor. 
It follows from the definition of $\mathsf{V}_{\tau}\ag{\mathsf{X}}$ and the \textit{square root theorem} \cite[e.g., see][\S2]{Gurtin1982} that
\begin{equation}
\mathsf{V}_{\tau}\ag{\mathsf{X}}
=\sum_{i\in \mathcal{I}}\sqrt{\varphi_{\tau i}\ag{\mathsf{X}}}\mathsf{n}_{\tau i}\ag{\mathsf{X}}\otimes \mathsf{n}_{\tau i}\ag{\mathsf{X}},
  \label{equ:le}
\end{equation}
where $\mathsf{n}_{\tau i}\ag{\mathsf{X}}\in \mathcal{M}_{3\times 1}\pr{\mathbb{R}}$ are the eigenvectors of $\mathsf{B}_{\tau}\ag{\mathsf{X}}$ such that
$\mathsf{n}_{\tau i}\ag{\mathsf{X}}\cdot \mathsf{n}_{\tau j}\ag{\mathsf{X}}=\delta_{ij}$ (which exist owing to the \textit{spectral theorem} \cite[\S2]{Gurtin1982}, since $\mathsf{B}_{\tau}\ag{\mathsf{X}}$ is symmetric), and
$\varphi_{\tau i}\ag{\mathsf{X}}$  are $\mathsf{B}_{\tau}\ag{\mathsf{X}}$'s eigenvalues corresponding to $\mathsf{n}_{\tau i}\ag{\mathsf{X}}$. 
The (spatial) logarithmic strain tensor $\ln\mathsf{V}_{\tau}\ag{\mathsf{X}}$ is defined to be the natural logarithm of  $\mathsf{V}_{\tau}\ag{\mathsf{X}}$. 
It can be shown that
\begin{equation}
  \ln\mathsf{V}_{\tau}\ag{\mathsf{X}} =\sum_{i\in \mathcal{I}}\pr{\ln\ag{\sqrt{\varphi_{\tau i}\ag{\mathsf{X}}}}}\mathsf{n}_{\tau i}\ag{\mathsf{X}}\otimes \mathsf{n}_{\tau i}\ag{\mathsf{X}}.
  \label{equ:le2}
\end{equation} 
It follows from \eqref{equ:le2} and the spectral theorem that $\ln \ag{\sqrt{\varphi_{\tau i}\ag{\mathsf{X}}}}$ are $\ln \mathsf{V}_{\tau}\ag{\mathsf{X}}$'s eigenvalues. 
Thus, the MPLS, $\phi_{\tau1}\ag{\mathsf{X}}$, is the maximum of the $\ln \ag{\sqrt{\varphi_{\tau i}\ag{\mathsf{X}}}}$, $i\in \mathcal{I}$. 

Writing $\mathsf{H}^{\star}_{\tau}\ag{\mathsf{X}}$ in our head model in terms of $U^{\mathcal{C}}_2\ag{\cdot, \cdot}$ using \eqref{eq:Hmatrix};
using that $\mathsf{H}^{\star}_{\tau}\ag{\mathsf{X}}$ and computing  $\mathsf{F}_{\tau}\ag{\mathsf{X}}$ using \eqref{eq:DeformationGradient};
computing the  $\mathsf{B}_{\tau}\ag{\mathsf{X}}$ corresponding to that $\mathsf{F}_{\tau}\ag{\mathsf{X}}$
using \eqref{eq:DefB}; computing that $\mathsf{B}_{\tau}\ag{\mathsf{X}}$'s eigenvalues and then computing the logarithm of the square root of the maximum of those eigenvalues we get that in our head model
\begin{equation}
\phi_{\tau1}\ag{\mathsf{X}}=\phi^{\mathcal{C}}_1\ag{\bar{r}\ag{\mathsf{X}}/r_0, \tau/\tau_1},
\label{eq:mpls}
\end{equation}
where
\begin{multline}
  \phi_1^{\mathcal{C}}\ag{\hat{r},\hat{\tau}}=\frac{1}{2}\ln \Vast[1+\frac{1}{2\hat{r}^2}\vast(\pr{\hat{U}^{\mathcal{C}}_2\ag{\hat{r},\hat{\tau}}}^2+\hat{r}^2\pr{\partial^{\pr{1,0}}\hat{U}^{\mathcal{C}}_2\ag{\hat{r},\hat{\tau}}}^2 \\
  +\abs{\hat{U}^{\mathcal{C}}_2\ag{\hat{r},\hat{\tau}}-\hat{r}\partial^{\pr{1,0}}\hat{U}^{\mathcal{C}}_2\ag{\hat{r},\hat{\tau}}}\sqrt{4\hat{r}^2+\pr{\hat{U}^{\mathcal{C}}_2\ag{\hat{r},\hat{\tau}}+\hat{r}\partial^{\pr{1,0}}\hat{U}^{\mathcal{C}}_2\ag{\hat{r},\hat{\tau}}}^2}
  \vast)\Vast].
\end{multline}


\subsubsection{Ninety fifth percentile maximum principal logarithimic strain, $\phi^{95}\ag{\cdot}$, and its maximum value $\phi^{95}_{\rm max}$}
\label{sec:MLPS95Def}
 The 95th percentile MPLS at the time instance $\tau$ is defined to be the MPLS value such that at time instance $\tau$ the MPLS values over  $95\%$ of the brain are smaller than it.
 We give a more precise definition below, which has encoded in it the details to compute it.

Let $\mathcal{G}:=\set{\mathcal{X} \in \mathcal{B}}{\breve{\sf X}_3 \circ \kappa_{\rm R} \ag{\mathcal{X}}=0}$ (for an illustration see Fig.~\ref{fig:notion}), and let $\phi_{\tau,\rm inf}$ and $\phi_{\tau,\rm sup}$ be the infimum and supremum of $\phi_{\tau 1}\ag{\cdot}$ over $\u{\sf{C}}\ag{\mathcal{G}}$, respectively, where
\[
\u{\sf{C}}\ag{\mathcal{G}}
=
\left\{
\mathsf{X}\in \mathcal{M}_{3 \times 1}(\mathbb{R})
~|~\sum_{i\in \mathcal{I}}
X_i \mathbf{E}_i \in
\boldsymbol{\kappa}_{\rm R}\ag{\mathcal{G}}
\right\}.
\]The $\phi$-sublevel set of $\phi_{\tau 1}\ag{\cdot }$  at the time instance $\tau$ is defined as
\[
S_{\tau}[\phi]=
\left\{
\mathsf{X}\in \u{\sf{C}}\ag{\mathcal{G}}~|~
\phi_{\tau 1}\ag{\mathsf{X}}
\le \phi
\right\}.
\]
We define the  map $m_{\tau}$,
\[
[\phi_{\tau,\rm inf}, \phi_{\tau,\rm sup}]
\ni
\phi\stackrel{m_{\tau} }{\to}
\frac{\text{meas}\ag{S_{\tau}\ag{\phi}}}{\text{meas}\ag{S_{\tau}\ag{\phi_{\rm max}}}}\in [0,1],
\]
where $\text{meas}\ag{\cdot}$ gives the Lebesgue measure of a set.
When $m_{\tau}$ has an inverse we define the $95^{\rm th}$ percentile MPLS  at the time instance $\tau$,  $\phi^{95}[\tau]$, to be $m_{\tau}^{-1}\ag{0.95}$.

The maximum 95th percentile MPLS, $\phi_{\rm max }^{95}$, is the supremum of the values attained by  $\phi^{95}\ag{\cdot}$ over all time. In our head model for arbitrary loadings it is not possible to determine the supremum  of $\phi^{95}\ag{\cdot}$.
This is because, since our head model does not include viscous effects and is non-linear, the dynamics in it may not reach a periodic state.
Therefore we take $\phi_{\rm max }^{95}$  in our head model to be the supremum of the values attained by  $\phi^{95}\ag{\cdot}$  only over a finite time interval.
Specifically, we take  $\phi_{\rm max }^{95}$ to be the supremum of the values taken by $\phi^{95}\ag{\cdot}$ over the time interval $[0,\tau_1+10 \tau_s)$.

\begin{figure}[H]
    \centering        \includegraphics[width=0.9\textwidth]{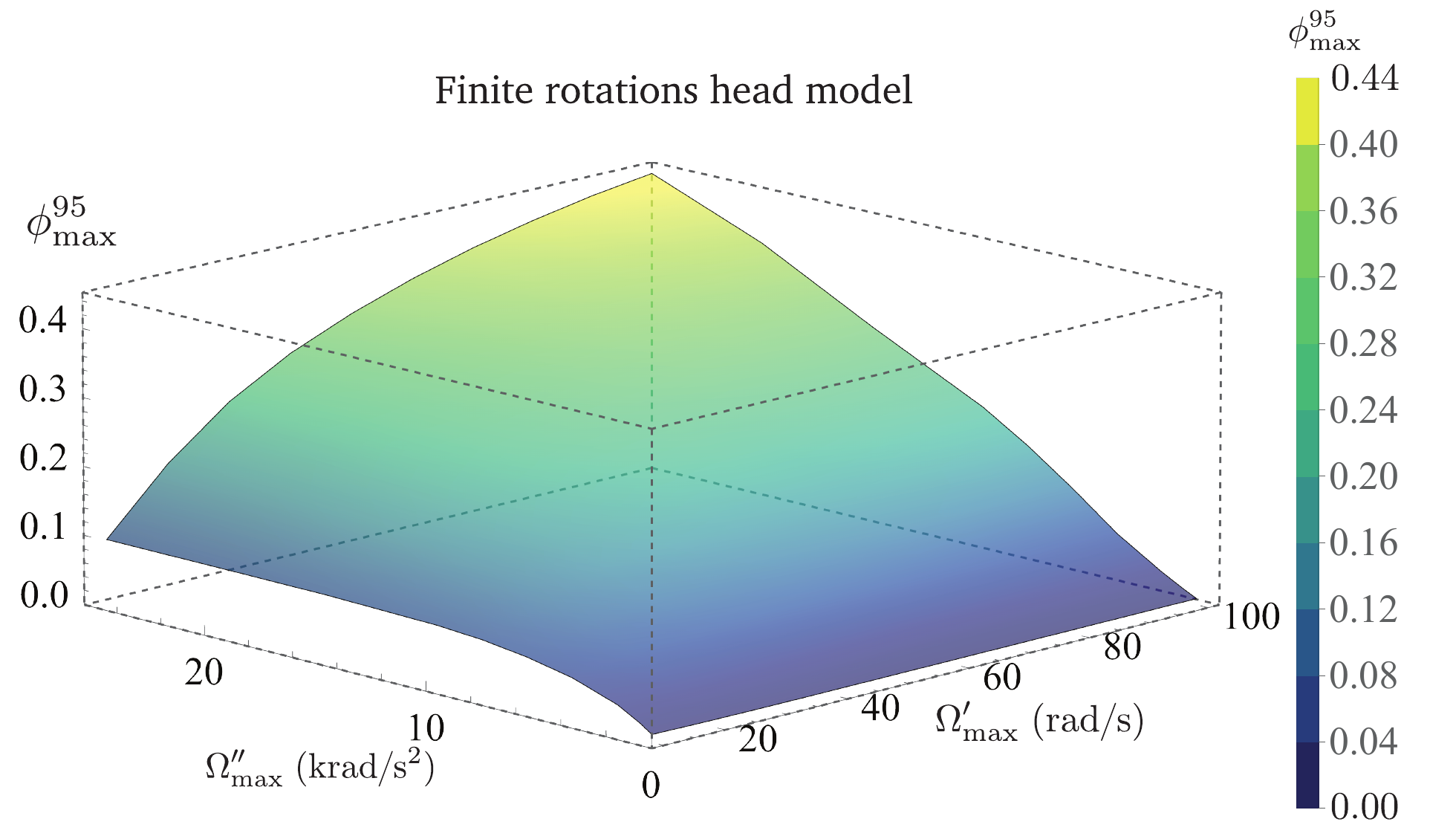}
    \caption{
    The maximum 95th percentile MPLS, $\phi^{95}_{\rm max}$, predicted by the finite rotations head model. The $\phi^{95}_{\rm max}$ values shown are from a family of motions; all of which correspond to the material and geometry parameters \eqref{eq:FEAMatProps}, and the loading function $\hat{\omega}\ag{\cdot}$ given in \eqref{equ:w2}. The family of motions were generate by varying $\pr{\tau_1, \tau_2}$ (resp. $\pr{\Omega'_{\rm max}, \Omega''_{\rm max}}$) from $\pr{33.14\times10^{-3},100\times10^{-3}}$ (resp. $\pr{10,0.5\times10^{3}}$) to $\pr{6.63\times10^{-3},10\times10^{-3}}$ (resp. $\pr{100,25\times10^{3}}$).
     The quantity $\phi^{95}_{\rm max}$ is discussed in \S\ref{sec:MLPS95Def}.
  }
    \label{fig:95mps}
\end{figure}


\subsubsection{Comparison with finite element solutions}
\label{sec:FEM}

As we mentioned in \S\ref{sec:femcompare}, we compared the ``maximum 95th percentile MPLS'' $\phi^{95}_{\rm max}$ values from the finite rotations head model with those from a related, finite element based, 2D head model \cite{Carlsen2021}. 
To make our head model approximate the finite element head model as closely as possible in our head model we take the non-dimensional head radius $r_0$, density $\rho_0$, and shear modulus $\mu$ to be
\begin{subequations}
\begin{align}
r_0&=0.06525,\\
\rho_0&=1040,\\
\mu&=34000.
\end{align}
\label{eq:FEAMatProps}
\end{subequations}
(It follows from \eqref{equ:ts} that the value of the parameter  $\tau_s$ for these geometry and material parameters is $0.0114$.) And we take $\hat{\omega}\ag{\cdot}$ to be
\begin{equation}
\hat{\tau}\mapsto \int_0^{\hat{\tau}}\hat{\psi}\ag{\xi}{\rm d}\xi,
 \label{equ:w2}
 \end{equation}
where
\begin{equation}
\hat{\psi}\ag{\hat{\tau}}:=
\begin{cases}
1.657~e^{1-\frac{1}{1-(2\hat{\tau}-1)^2}}, & 0\leq \hat{\tau}\leq 1,\\
0, & \hat{\tau}> 1.
\end{cases}
\label{equ:wpre2}
\end{equation}
The function $\hat{\omega}'\ag{\cdot}$ corresponding to the $\hat{\omega}\ag{\cdot}$ given in \eqref{equ:w2} is the function $\hat{\psi}\ag{\cdot}$ given in \eqref{equ:wpre2}.
The graph of the functions \eqref{equ:w2} and \eqref{equ:wpre2} are sketched in Figs.~\ref{fig:bump}(e) and (f), respectively.
The loading in our and the finite element head model  depends not only on $\hat{\omega}\ag{\cdot}$ (and hence $\hat{\omega}'\ag{\cdot}$) but also on $\tau_1$ and $\tau_2$.

 It can be shown that for the choice of \eqref{equ:w2} for $\hat{\omega}\ag{\cdot}$ the parameters $(\tau_1,\tau_2)$ depend on $(\Omega'_{\rm max},\Omega''_{\rm max})$, which we will define shortly, as
\begin{subequations}
\begin{equation}
\Omega'_{\rm max}=\frac{1}{\tau_2},
\label{equ:maxvel}
\end{equation}
and
\begin{equation}
\Omega''_{\rm max}=\frac{1.657}{\tau_1\tau_2}.
\label{equ:maxacce}
\end{equation}
\label{equ:maxvelacce}
\end{subequations}
 The parameters  $\Omega'_{\rm max}$ and $\Omega''_{\rm max}$ are, respectively, the maximum values of $\Omega'\ag{\cdot}$ and $\Omega''\ag{\cdot}$.

\begin{figure}[H]
    \centering
        \includegraphics[width=0.9\textwidth]{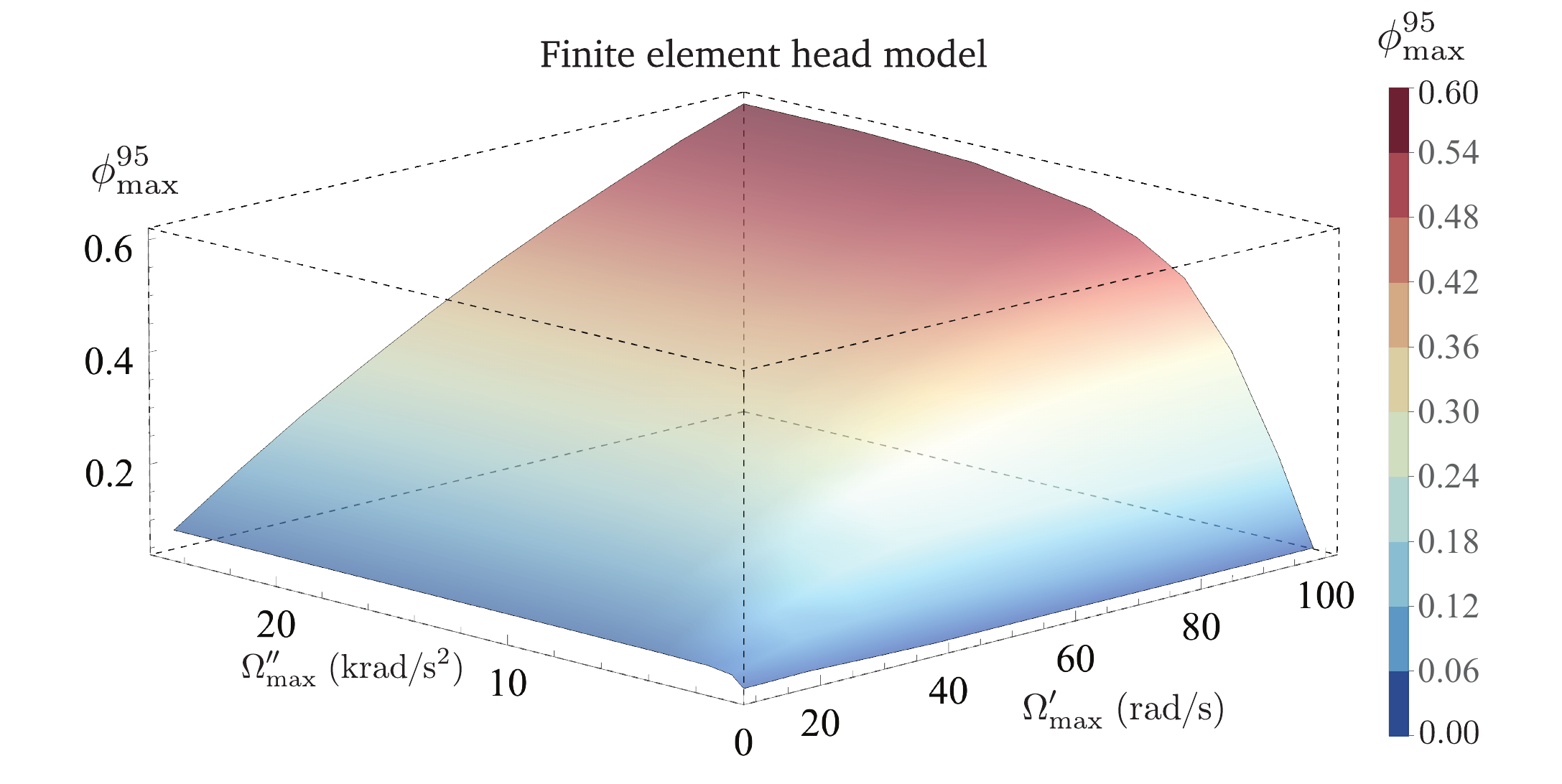}
    \caption{
    The maximum 95th percentile MPLS, $\phi^{95}_{\rm max}$, predicted by the  2D finite element head model presented in \cite{Carlsen2021} for a family of loadings. Those loadings are same as the loadings used to generate the family of motions  whose $\phi^{95}_{\rm max}$ values are shown in Fig.~\ref{fig:95mps}.  Some details of how the 2D finite element head model was created are as follows.
    The model was generated by processing magnetic resonance images (MRI) and diffusion tensor images (DTI) of an adult subject.
    It included all major anatomical structures, such as the skull, white matter, gray matter, cerebral spinal fluid, ventricles, bridging veins, falx cerebri, and tentorium cerebelli.
    Different 2D finite element head models were created by taking cross-sections of the head geometry along the sagittal, coronal, and axial planes.
    Plane strain conditions were assumed in each of those 2D finite element models.
    The $\phi^{95}_{\rm max}$ values in this figure and Fig.~\ref{fig:95mpsrikacom} are from the 2D coronal finite element head model.
    The brain tissue in the 2D coronal finite element head model was modeled as an anisotropic hyper-viscoelastic material using the Holzapfel-Gasser-Ogden strain energy function as described in detail in Wright \emph{et al.} \cite{wright2013}.
    The cerebral spinal fluid in it was modeled using a Mie-Gruneisen equation of state.
    The material interfaces in the model were tied so that no tangential sliding or normal separation could occur.
    }
    \label{fig:95mpsrika}
\end{figure}

\begin{figure}[H]
    \centering
        \includegraphics[width=\textwidth]{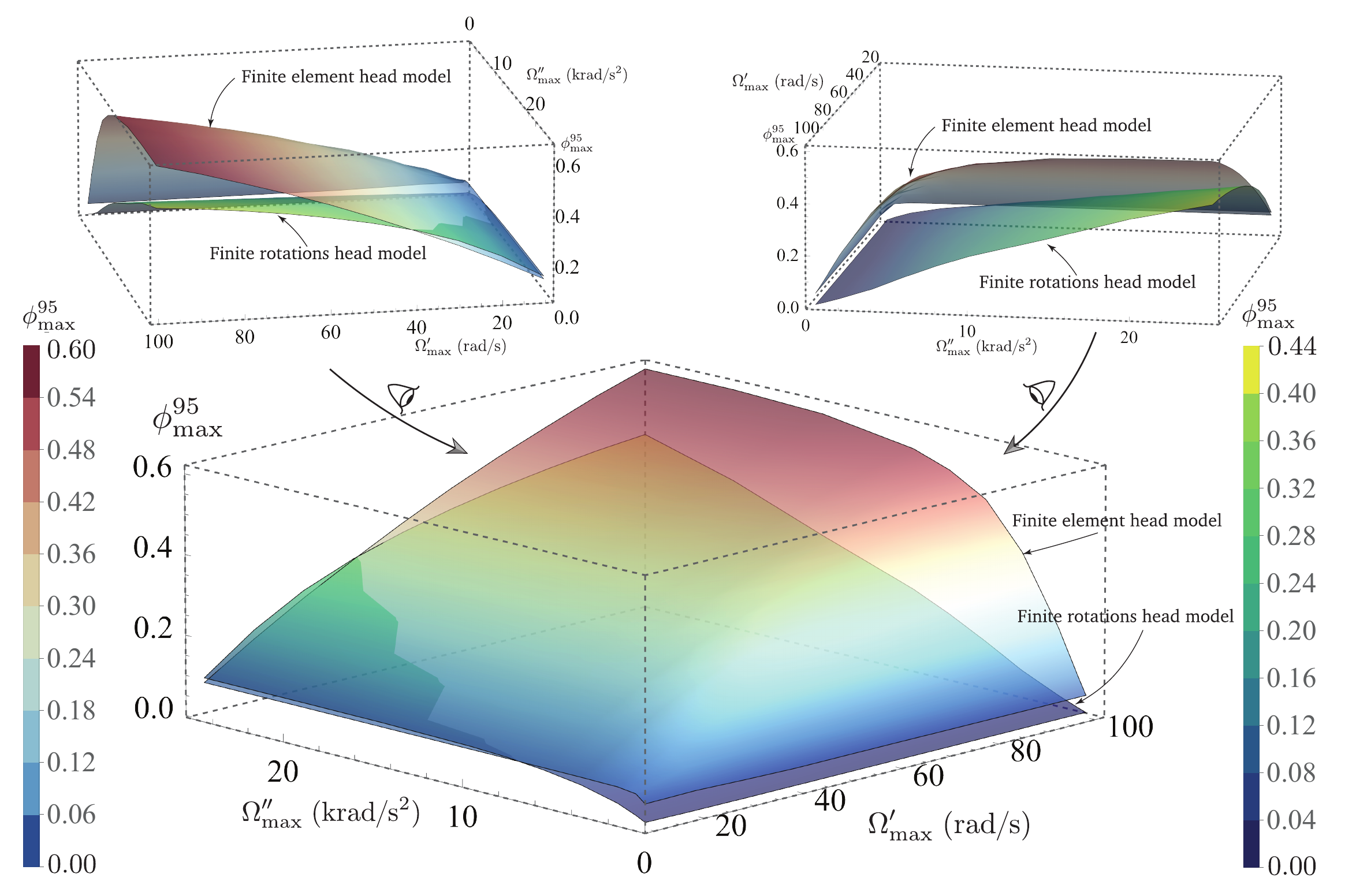}
    \caption{Comparison of the maximum 95th percentile MPLS, $\phi^{95}_{\rm max}$, values from the  finite rotations head model and the finite element head model.
    The $\phi^{95}_{\rm max}$ values for the finite rotations head model are the ones from  Fig.~\ref{fig:95mps}.
    The $\phi^{95}_{\rm max}$ values for  the finite element head model are the ones from  Fig.~\ref{fig:95mpsrika}.
        The insets in the figure show different views of the 3D plot shown in this figure.
  }
    \label{fig:95mpsrikacom}
\end{figure}

We considered a range of values for $(\Omega'_{\rm max},\Omega''_{\rm max})$. For each of that  $(\Omega'_{\rm max},\Omega''_{\rm max})$ values we computed the corresponding $(\tau_1, \tau_2)$ value using \eqref{equ:maxvelacce}.
For that $(\tau_1, \tau_2)$ value and the chosen $r_0$, $\tau_s$, $\hat{\omega}\ag{\cdot}$, $\hat{\omega}'\ag{\cdot}$ we computed  $\hat{U}^{\mathcal{C}}_2\ag{\cdot, \cdot}$ by following the first six steps given in procedure \ref{algo:ThreeTermdisp}.
Using that $\hat{U}^{\mathcal{C}}_2\ag{\cdot, \cdot}$ and \eqref{eq:mpls} we then computed the maximum 95th percentile MPLS, $\phi^{95}_{\rm max}$.
We plot the computed $\phi^{95}_{\rm max}$ as a function of $(\Omega'_{\rm max},\Omega''_{\rm max})$ in Fig.~\ref{fig:95mps}.
We plot the $\phi^{95}_{\rm max}$ from the finite element calculations as a function of $(\Omega'_{\rm max},\Omega''_{\rm max})$ in Fig.~\ref{fig:95mpsrika}.
In Fig.~\ref{fig:95mpsrikacom} we show both the $\phi^{95}_{\rm max}$ from our finite rotations head model as well as that from the finite element head model.
The insets in Fig.~\ref{fig:95mpsrikacom} show different views of the 3D plot shown in Fig.~\ref{fig:95mpsrikacom}.

As can be seen from Fig.~\ref{fig:95mpsrikacom}, the $\phi^{95}_{\rm max}$ values from our head model are comparable to those from the finite element head model.
For example, for $\pr{\Omega'_{\rm max},\Omega''_{\rm max}}=\pr{100, 25000}$ our model predicts a value of $0.44$  for $\phi^{95}_{\rm max}$, whereas the finite element head model predicts a value of $0.6$.
The maximum difference in the predicted values for $\phi^{95}_{\rm max}$ from  our finite rotations head model and the finite element head model is 0.36 (finite rotations head  model $0.17$; finite element head model  $0.53$). This occurs when $\pr{\Omega'_{\rm max},\Omega''_{\rm max}}=\pr{100, 8000}$.
The dependence of $\phi^{95}_{\rm max}$ on $\pr{\Omega'_{\rm max},\Omega''_{\rm max}}$ in our model is qualitatively similar to the dependence of $\phi^{95}_{\rm max}$ on $\pr{\Omega'_{\rm max},\Omega''_{\rm max}}$ in the finite element head model.
As we stated in \S\ref{sec:femcompare} we find it remarkable that  $\phi^{95}_{\rm max}$ values from our head model and the finite element head model are as close as they are in Fig.~\ref{fig:95mpsrikacom} considering that the finite element head model is far more sophisticated then our head model. We give some of the details of the finite element head model in the caption of Fig.~\ref{fig:95mpsrika}.





\subsection{Comparison with previous idealized continuum mechanics based head models}
\label{sec:smallrotation}

 As we mentioned in the introduction \S\ref{sec:intro}, all the previous idealized continuum mechanics based head models that we surveyed take the head's rotations to be small. For that reason, we refer to them also as the \emph{small rotations head models}.

\subsubsection{Comparison of the governing equations}
\label{sec:comgover}
Recall that we introduced the tangential component of displacement $u^{\mathcal{C}}_2\ag{\cdot, \cdot}$ in \eqref{eq:uintermsofCylComponents}. Introducing the scaled tangential  displacement component  $\hat{u}^{\mathcal{C}}_2\ag{\cdot,\cdot}: \ag{0,1}\times\mathbb{R}_{ \ge 0} \rightarrow \mathbb{R}$ as
\begin{equation}
\hat{u}^{\mathcal{C}}_2\ag{\hat{r},\hat{\tau}}=u^{\mathcal{C}}_2\ag{\hat{r}r_0,\hat{\tau}\tau_1}/r_0,
\label{equ:scaleddispu2}
\end{equation}
we can write the initial boundary value problem (IBVP) \eqref{equ:scalegoverninge2all} in terms of $\hat{u}^{\mathcal{C}}_2\ag{\cdot, \cdot}$ as follows.
\begin{subequations}
The governing partial differential equation (PDE) \eqref{equ:scalegoverninge2} in terms of $\hat{u}^{\mathcal{C}}_2\ag{\cdot, \cdot}$ reads
\begin{equation}
\begin{split}
 \partial^{\pr{2,0}} \hat{u}^{\mathcal{C}}_2\ag{\hat{r},\hat{\tau}}+\frac{\partial^{\pr{1,0}} \hat{u}^{\mathcal{C}}_2 \ag{\hat{r},\hat{\tau}}}{\hat{r}}-\frac{\hat{u}^{\mathcal{C}}_2\ag{\hat{r},\hat{\tau}}}{\hat{r}^2}=&\frac{\tau_s^2\partial^{\pr{0,2}} \hat{u}^{\mathcal{C}}_2\ag{\hat{r},\hat{\tau}}}{\tau_1^2} \\
  &+\frac{\tau_s^2\tan \circ~\hat{\Omega}\ag{\hat{\tau}}\sec\circ~\hat{\Omega}\ag{\hat{\tau}}}{2\tau_2^2}\Bigg(-3\hat{r}+\hat{r}\cos\circ~2\hat{\Omega}\ag{\hat{\tau}}\\
&~~~~~+4\sin\circ~\hat{\Omega}\ag{\hat{\tau}}\hat{u}^{\mathcal{C}}_2\ag{\hat{r},\hat{\tau}}\Bigg)\hat{\omega}\ag{\hat{\tau}}^2\\
  &+\frac{\tau_s^2}{\tau_1\tau_2}\tan\circ~\hat{\Omega}\ag{\hat{\tau}}\pr{-\hat{r}\sin\circ~\hat{\Omega}\ag{\hat{\tau}}+\hat{u}^{\mathcal{C}}_2\ag{\hat{r},\hat{\tau}}}\hat{\omega}'\ag{\hat{\tau}}\\
&+\frac{2\tau_s^2}{\tau_1\tau_2}\tan\circ~\hat{\Omega}\ag{\hat{\tau}}\partial^{\pr{0,1}} \hat{u}^{\mathcal{C}}_2\ag{\hat{r},\hat{\tau}}\hat{\omega}\ag{\hat{\tau}},
\label{eq:scaledsgoverning1}
\end{split}
\end{equation}
where $\hat{\Omega}\ag{\cdot}:\mathbb{R}_{ \ge 0} \rightarrow \mathbb{R}$, is defined as
\begin{equation}
\hat{\Omega}\ag{\hat{\tau}}=\Omega\ag{\hat{\tau}\tau_1}.
\label{equ:scaledOmega}
\end{equation}
The boundary condition \eqref{equ:scalebc22} reads
\begin{equation}
\hat{u}^{\mathcal{C}}_2\ag{\hat{r}=1,\hat{\tau}}=\sin \circ~\hat{\Omega}\ag{\hat{\tau}},
\label{equ:scaleduBCs}
\end{equation}
and the initial conditions \eqref{equ:scaleic21} and \eqref{equ:scaleic22} read
\doubleequation[equ:scaleuic21,equ:scaleuic22]{
\hat{u}^{\mathcal{C}}_2\ag{\hat{r}, \hat{\tau}=0}=0,}
{
\partial^{\pr{0,1}} \hat{u}^{\mathcal{C}}_2\ag{\hat{r},\hat{\tau}=0}=0.}
\label{eq:BVPhatu2}
\end{subequations}

When the the skull's rotations are small, the governing PDE of the IBVP \eqref{eq:scaledsgoverning1} reduces to
\begin{subequations}

\begin{equation}
 \partial^{\pr{2,0}} \hat{u}^{\mathcal{C}}_2\ag{\hat{r},\hat{\tau}}+\frac{\partial^{\pr{1,0}} \hat{u}^{\mathcal{C}}_2 \ag{\hat{r},\hat{\tau}}}{\hat{r}}-\frac{\hat{u}^{\mathcal{C}}_2\ag{\hat{r},\hat{\tau}}}{\hat{r}^2}=\frac{\tau_s^2\partial^{\pr{0,2}} \hat{u}^{\mathcal{C}}_2\ag{\hat{r},\hat{\tau}}}{\tau_1^2},
  \label{eq:sgoverning}
\end{equation}
and the boundary condition \eqref{equ:scaleduBCs} to
\begin{equation}
\hat{u}^{\mathcal{C}}_2\ag{\hat{r}=1,\hat{\tau}}=\hat{\Omega}\ag{\hat{\tau}}.
\label{equ:scaledsuBCs}
\end{equation}
\label{eq:u2CBVPLinearized}
\end{subequations}
The initial conditions \eqref{equ:scaleuic21}, \eqref{equ:scaleuic22} remain the same. In arriving at \eqref{eq:u2CBVPLinearized} we ignored terms of $o\pr{\hat{\Omega}}$.

Most of the 2D ICM based head models presented to date include viscous effects.
To our knowledge all the 2D ICM based head models reported to date  reduce to the IBVP \eqref{eq:u2CBVPLinearized} on ignoring any viscous effects in them. Some of such 2D head models are the ones reported by Ljung \cite{Ljung1975}, Margulies and Thibault \cite{Margulies1989}, Massouros \emph{et al.} \cite{Massouros2014}, and Massouros \cite{Massouros2005}.

Bayly \emph{et al.}\cite{Bayly2008} studied the dynamics of a viscoelastic cylinder where the loading is prescribed through the motion of the cylinder's boundary.
On ignoring the viscous effects the problem studied in \cite{Bayly2008} too reduces to the problem \eqref{eq:u2CBVPLinearized}.

\subsubsection{Comparison of a quantitative prediction}
\label{sec:comMPLS}
In order to check how quantitatively different our finite rotations head model is to the small rotations head models we again compared predictions for   $\phi^{95}_{\rm max}$.
For this comparison we again took the material and geometry parameters to be those given by \eqref{eq:FEAMatProps}, and the loading function $\hat{\omega}\ag{\cdot}$ to be the one given in \eqref{equ:w2}; and considered the same range of $(\tau_1, \tau_2)$ (equivalently $(\Omega'_{\rm max},\Omega''_{\rm max})$) values that we used for comparing our head model with the finite element head model in \S\ref{sec:FEM}.

The procedure for computing $\phi^{95}_{\rm max}$ in the small rotations head models is almost the same as the procedure given in \S\ref{sec:MLPSDef} and \S\ref{sec:MLPS95Def} for computing $\phi^{95}_{\rm max}$ in our finite rotations head model. The primary difference is that in the small rotations head models the MPLS is
\begin{equation}
\frac{1}{2}\ln\pr{1+\frac{\abs{u^{\mathcal{C}}_2\ag{\bar{r}\ag{\sf X},\tau}-\bar{r}\ag{\sf X}\partial^{\pr{1,0}}u^{\mathcal{C}}_2\ag{\bar{r}\ag{\sf X},\tau}}}{\bar{r}\ag{\sf X}}},
\label{equ:95MPLSsmallrotation}
\end{equation}
where $u^{\mathcal{C}}_2\ag{\cdot, \cdot}$ is the solution of \eqref{eq:sgoverning} subjected to the boundary condition \eqref{equ:scaledsuBCs}, and the initial  conditions \eqref{equ:scaleuic21}--\eqref{equ:scaleuic22}. We solve the initial boundary value problem for $u^{\mathcal{C}}_2\ag{\cdot, \cdot}$ in the small rotations head models numerically.

\begin{figure}[H]
    \centering
        \includegraphics[width=0.95\textwidth]{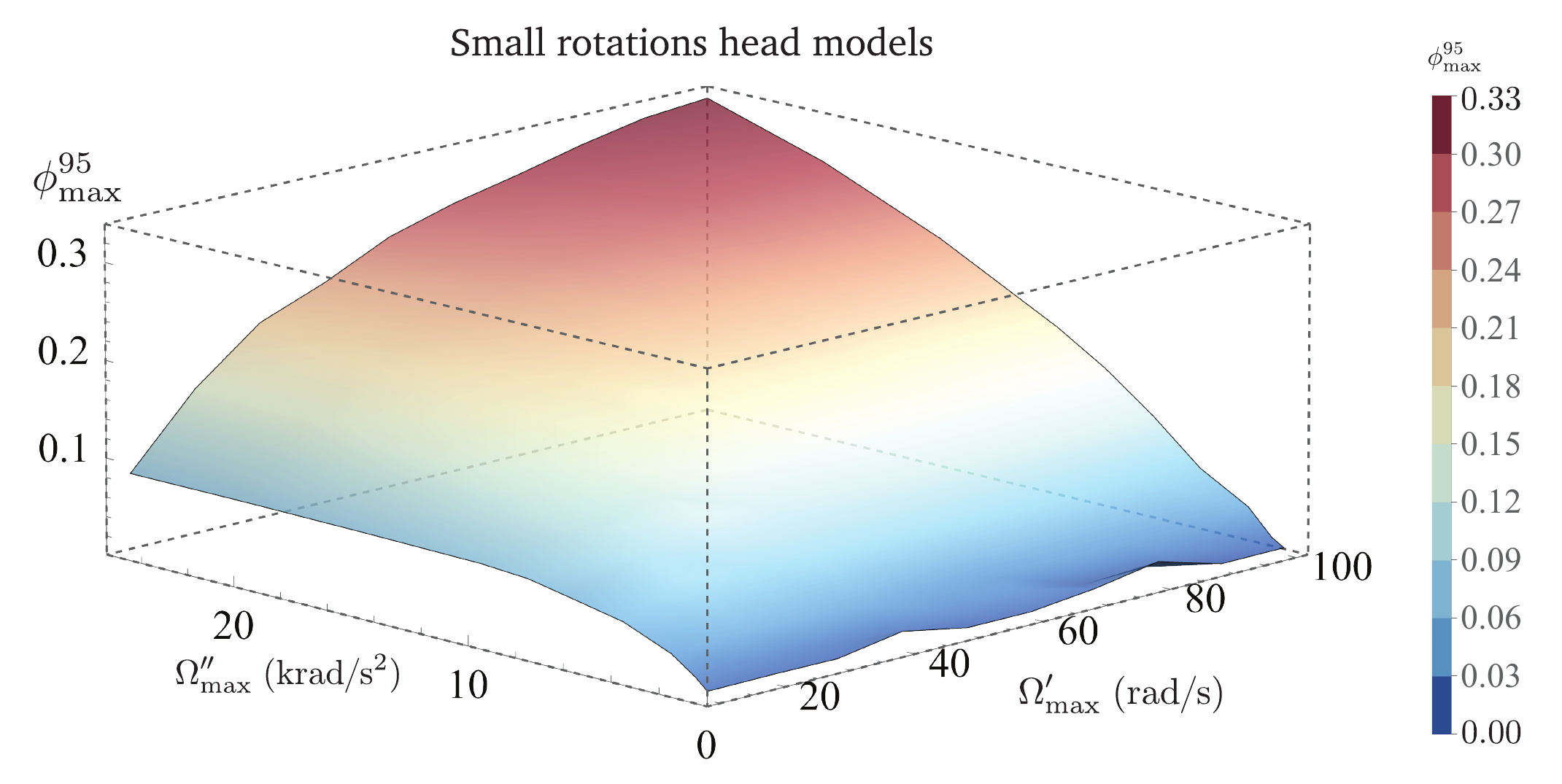}
    \caption{The maximum 95th percentile MPLS, $\phi^{95}_{\rm max}$, predicted by the small rotations head model (see \S\ref{sec:comgover} for details). The $\phi^{95}_{\rm max}$ values shown are from a family of motions,  all which correspond to the same  material and geometry properties,  and loading function; but a range of $(\tau_1, \tau_2)$ (resp. $\pr{\Omega'_{\rm max}, \Omega''_{\rm max}}$) values. The  material and geometry properties, the loading function, and the range of  $\pr{\tau_1, \tau_2}$ values are the same as those used for generating the motions in Fig.~\ref{fig:95mps}. The $\phi^{95}_{\rm max}$ values were computed by generating the family of motions by solving the equations given in \S\ref{sec:comgover} and using \eqref{equ:95MPLSsmallrotation}.
  }
    \label{fig:95mpssurveyedmodel1}
\end{figure}

\begin{figure}[H]
    \centering
        \includegraphics[width=\textwidth]{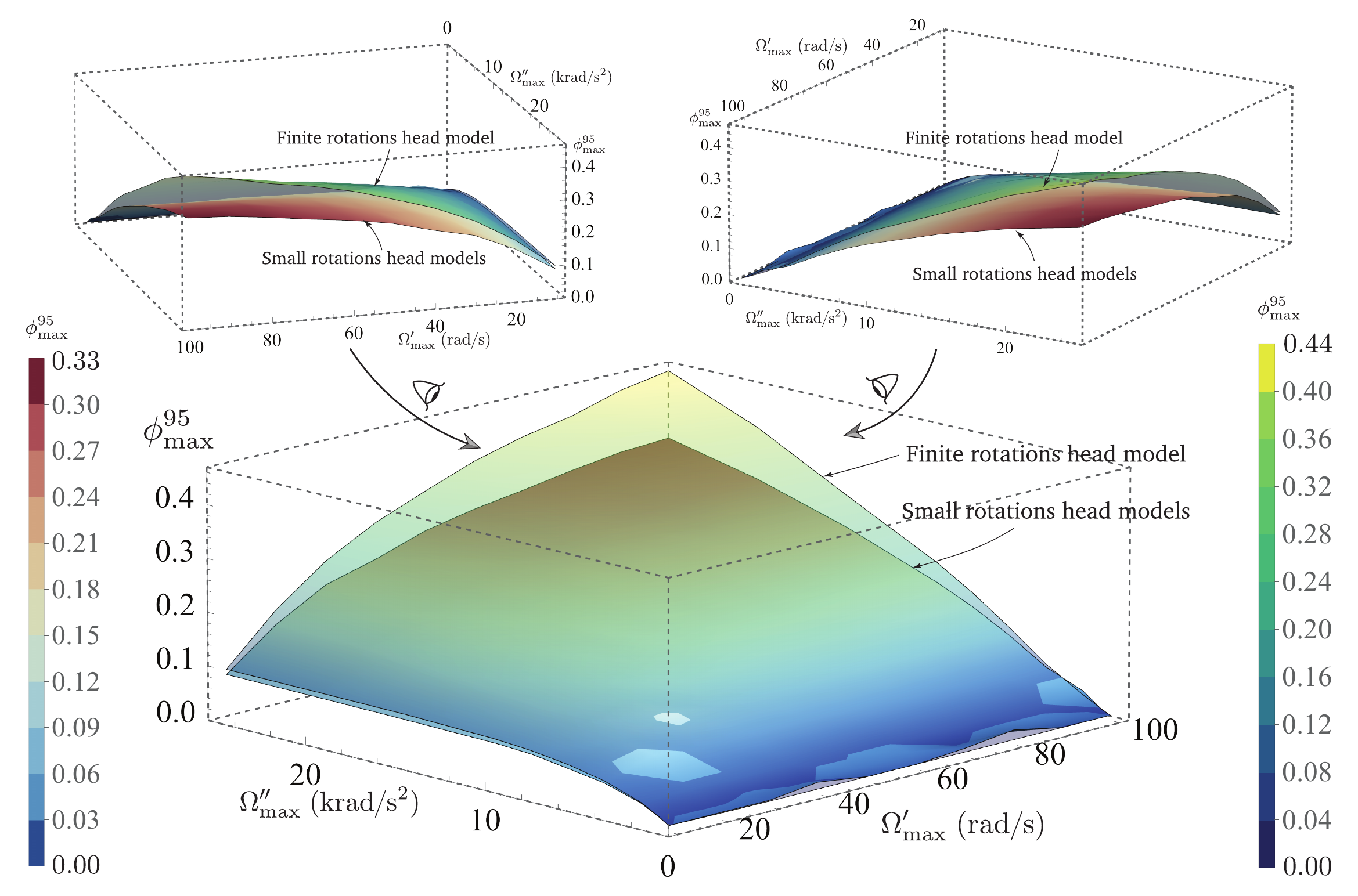}
    \caption{Comparison of the maximum 95th percentile MPLS, $\phi^{95}_{\rm max}$, values from the  finite rotations head model and the small rotations head model.
The $\phi^{95}_{\rm max}$ values for the finite rotations head model are the ones from  Fig.~\ref{fig:95mps}.
The $\phi^{95}_{\rm max}$ values for the small rotations head model are the ones from Fig.~\ref{fig:95mpssurveyedmodel1}.
The insets in the figure show different views of the 3D plot shown in this figure.
  }
    \label{fig:95mpssurveyedmodel2}
\end{figure}

We plot $\phi^{95}_{\rm max}$ from the small rotations head models as a function of $(\Omega'_{\rm max},\Omega''_{\rm max})$ in Fig.~\ref{fig:95mpssurveyedmodel1}.
In Fig.~\ref{fig:95mpssurveyedmodel2} we show both the $\phi^{95}_{\rm max}$ from the small rotations head models as well as that from our finite rotations head model.
The insets in Fig.~\ref{fig:95mpssurveyedmodel2} show different views of the 3D plot shown in Fig.~\ref{fig:95mpssurveyedmodel2}.

As can be seen in Fig.~\ref{fig:95mpssurveyedmodel2} except for when $\Omega'_{\rm max}$ and $\Omega''_{\rm max}$ are small the predictions from the small rotations head models are quite different from those from the finite rotations head model. For example, for $\pr{\Omega'_{\rm max},\Omega''_{\rm max}}=\pr{100, 25000}$ the small rotations head model predicts a value of $0.33$ for $\phi^{95}_{\rm max}$, whereas our model predicts a value of $0.44$.

\section{Concluding remarks}
\label{sec:con}

\begin{enumerate}

\item In deriving the finite rotations head model, we have assumed the brain to be homogeneous and isotropic; to undergo small strains; have no viscous effects; and even more dramatically, to undergo 2D deformation. Considering these type of simplifications, we find it remarkable that the ICM head models are  capable of providing the kind of first order estimates for the peak strains and strain rates as those shown in Figs.~\ref{fig:95mps}, \ref{fig:95mpsrika}, \ref{fig:95mpsrikacom}, \ref{fig:95mpssurveyedmodel1}, and \ref{fig:95mpssurveyedmodel2}.

\item We plan on incorporating viscous effects and considering the 3D nature of the brain's deformation shortly in the future.   However, despite those augmentations, it is likely that the estimates provided by computational mechanics (CM) based head models will be more accurate than those provided by our, or any other idealized continuum mechanics (ICM) based, head model.  As we mentioned in the introduction, the primary advantage of ICM based head models compared to CM based models is that they're much easier and faster to apply for assessing the injury risk of a mechanically traumatic event.

    \item    Given the head geometry, brain material property details, and a quantitative representation of the loading, procedure \ref{algo:ThreeTermdisp} can be used for determining the displacements in the finite rotations head model.  Strains and strain rates can then be computed from the displacements by using \eqref{eq:strain} and \eqref{eq:strainrate}, respectively.   Though this procedure is far simpler than that involved in using a CM based model, health care and medical professionals interested in mTBI may still find it difficult to apply our procedure. In order to make the finite rotations head model easy to apply, especially by non-engineers, we have built a web application that automatically solves the finite rotations head model and provides the displacements, strains, and strain rates  for a given loading input.  This web app can be accessed at \url{http://18.233.10.106:8501/}. 


\end{enumerate}




\section*{Acknowledgements}
The authors gratefully acknowledge support from the Panther Program, Tiger Program, and the Office of Naval Research (Dr. Timothy Bentley) under grants N000142112044 and N000142112054.
\section*{Declaration of Competing Interest}
The authors declare that they have no known competing financial interests or personal relationships that could have appeared to influence the work reported in this paper.

\appendix


\section{Derivation of $U^{\mathcal{C}}_1=0$}
\label{sec:incom}
Following the assumption that $\mathcal{B}$ is incompressible (assumption \textit{A.i.c} in \S\ref{sec:ProblemStatement}), we have that
\begin{equation}
{\sf Det}\ag{{\sf F}_{\tau}\ag{\sf X}}=1,
\label{eq:IncompressibilityConstraint}
\end{equation}
where ${\sf Det}$ is determinant operator.
Using \eqref{eq:deformgradient1} and substituting $\mathsf{F}_{\tau}\ag{\sf X}$ as ${\sf Q}\ag{\tau}{\sf F}^{\star}_{\tau}\ag{\sf X}$ in \eqref{eq:IncompressibilityConstraint}; in the resulting equation, using \eqref{eq:deformgradient} and substituting    ${\sf F}^{\star}_{\tau}\ag{\sf X}$ as ${\sf I}+{\sf H}^{\star}_{\tau}\ag{\sf X}$; and then noting that ${\sf Det}\ag{\mathsf{Q}\ag{\tau}}=1$, since $\mathsf{Q}\ag{\tau} \in SO(3)$, we get that
\begin{equation}
{\sf Det}\ag{{\sf I}+{\sf H}^{\star}_{\tau}\ag{\sf X}}=1.
  \label{equ:DetH1}
\end{equation}
From our assumption that the displacements and deformations of the brain w.r.t. the skull are uniformly small (assumption \textit{A.iii} in \S\ref{sec:ProblemStatement} ) and the identity that for any $\mathsf{H}\in \mathcal{M}_{3\times 3}\pr{\mathbb{R}}$ as $\epsilon \to 0$
\begin{equation}
\textsf{Det}\ag{\mathsf{I}+\epsilon \mathsf{H}}=1+\epsilon \textsf{Tr}\ag{\mathsf{H}}+o\pr{\epsilon},
  \label{equ:DetH2}
\end{equation}
we get that up to terms of $o\pr{\mathsf{H}^{\star}_{\tau}\ag{\mathsf{X}}}$,
\begin{equation}
{\sf Tr}\ag{{\sf H}^{\star}_{\tau}\ag{\sf X}}=0.
  \label{equ:DetH3}
\end{equation}
Writing ${\sf H}^{\star}_{\tau}$ in \eqref{equ:DetH3} in terms of $\pr{U^{\mathcal{C}}_i}_{i \in \mathcal{I}}$ using \eqref{eq:StrainDisplacement}, \eqref{eq:Utaui}, and \eqref{equ:disU1} we get that
\begin{equation}
  r \partial^{\pr{1,0}}U^{\mathcal{C}}_{1}\ag{r,\tau}+
  U^{\mathcal{C}}_{1}\ag{r,\tau}=0,
  \label{eq:DivUzero}
\end{equation}
for $r\in (0,r_0)$.
It follows from our assumptions \textit{A.i.e}  (skull is a rigid solid), and \textit{A.vi} (brain is rigidly connected to the skull) that
\begin{equation}
U^{\mathcal{C}}_1\ag{r=r_0,\tau}=0.
\label{eq:bc2A}
\end{equation}
By solving \eqref{eq:DivUzero} with the boundary condition \eqref{eq:bc2A} it can be shown that
\begin{equation}
U^{\mathcal{C}}_1\ag{r,\tau}=0,
\label{equ:u1czero}
\end{equation}
for all admissible $r$ and $\tau$.

\bibliography{refer}

\begin{thebibliography}{10}
\bibitem{li2021subject}
X.~Li, Subject-specific head model generation by mesh morphing: A
  personalization framework and its applications, Frontiers in Bioengineering
  and Biotechnology 9.
\end{thebibliography}

\end{document}